\documentstyle[12pt,epsf,rotate]{article}

\def\spose#1{\hbox to 0pt{#1\hss}}
\def\lsim{\mathrel{\spose{\lower 3pt\hbox{$\mathchar"218$}}
 \raise 2.0pt\hbox{$\mathchar"13C$}}}
\def\gsim{\mathrel{\spose{\lower 3pt\hbox{$\mathchar"218$}}
 \raise 2.0pt\hbox{$\mathchar"13E$}}}

\begin{document}

\begin{titlepage}

\begin{flushright}
CERN-TH/98-60\\
hep-ph/9802433
\end{flushright}

\vspace{1.4cm}
\begin{center}
\boldmath
\large\bf
Rescattering and Electroweak Penguin Effects in\\ 
\vspace{0.2truecm}
Strategies to Constrain and Determine the\\ 
\vspace{0.2truecm}
CKM Angle $\gamma$ from $B\to\pi K$ Decays
\unboldmath
\end{center}

\vspace{1cm}
\begin{center}
Robert Fleischer\\[0.1cm]
{\sl Theory Division, CERN, CH-1211 Geneva 23, Switzerland}
\end{center}

\vspace{1cm}
\begin{abstract}
\vspace{0.2cm}\noindent
A general parametrization of the $B^+\to\pi^+ K^0$ and $B_d^0\to\pi^- K^+$ 
decay amplitudes is presented. It relies only on the isospin symmetry of 
strong interactions and the phase structure of the Standard Model and 
involves no approximations. In particular, this parametrization takes into 
account both rescattering and electroweak penguin effects, which limit the 
theoretical accuracy of bounds on $\gamma$ arising from the combined
$B^\pm\to\pi^\pm K$, $B_d\to\pi^\mp K^\pm$ branching ratios. Generalized 
bounds making also use of the CP asymmetry in the latter decay are derived, 
and their sensitivity to possible rescattering and electroweak penguin 
effects is investigated. It is pointed out that experimental data on 
$B^\pm\to K^\pm K$ allow us to include rescattering processes in these 
bounds completely, and an improved theoretical treatment of electroweak  
penguins is presented. It is argued that rescattering effects may enhance the 
combined $B^\pm\to K^\pm K$ branching ratio by a factor of ${\cal O}(10)$ 
to the $10^{-5}$ level, and that they may be responsible for the small 
present central value of the ratio of the combined $B_d\to\pi^\mp K^\pm$ and 
$B^\pm\to\pi^\pm K$ branching ratios, which has recently been reported by 
the CLEO collaboration and, if confirmed, would exclude values of $\gamma$ 
within a large region around $90^\circ$.

\end{abstract}

\vfill
\noindent
CERN-TH/98-60\\
February 1998

\end{titlepage}

\thispagestyle{empty}
\vbox{}
\newpage
 
\setcounter{page}{1}

\section{Introduction}\label{intro}
The decays $B^+\to\pi^+K^0$, $B^0_d\to\pi^-K^+$ and their charge-conjugates 
offer a way \cite{PAPIII}--\cite{babar} to obtain experimental information 
on the angle $\gamma$ of the usual, non-squashed, unitarity triangle 
\cite{ut} of the Cabibbo--Kobayashi--Maskawa matrix (CKM matrix) 
\cite{ckm} at future $B$ factories. Recently, these decays have been observed 
by the CLEO collaboration~\cite{cleo} and experimental data are now starting 
to become available. So far only results for the combined branching ratios: 
\begin{eqnarray}
\mbox{BR}(B^\pm\to\pi^\pm K)&\equiv&\frac{1}{2}\left[\mbox{BR}(B^+\to\pi^+K^0)
+\mbox{BR}(B^-\to\pi^-\overline{K^0})\right]\label{BR-char}\\
\mbox{BR}(B_d\to\pi^\mp K^\pm)&\equiv&\frac{1}{2}
\left[\mbox{BR}(B^0_d\to\pi^-K^+)
+\mbox{BR}(\overline{B^0_d}\to\pi^+K^-)\right]\label{BR-neut}
\end{eqnarray}
have been published, with large experimental uncertainties \cite{cleo}:
\begin{eqnarray}
\mbox{BR}(B^\pm\to\pi^\pm K)&=&\left(2.3^{+1.1}_{-1.0}\pm0.3\pm0.2\right)
\times 10^{-5}\label{BR-charres}\\
\mbox{BR}(B_d\to\pi^\mp K^\pm)&=&\left(1.5^{+0.5}_{-0.4}\pm0.1\pm0.1
\right)\times10^{-5}\,.\label{BR-neutres}
\end{eqnarray}
Consequently, it is not yet possible to determine $\gamma$ as was 
proposed in \cite{PAPIII,groro}. However, as was pointed out in \cite{fm2}, 
these combined branching ratios imply {\it bounds} on $\gamma$, which 
are of the form
\begin{equation}\label{gamma-bound1}
0^\circ\leq\gamma\leq\gamma_0\quad\lor\quad180^\circ-\gamma_0\leq\gamma
\leq180^\circ,
\end{equation}
and are hence complementary to the presently allowed range $41^\circ
\mathrel{\hbox{\rlap{\hbox{\lower4pt\hbox{$\sim$}}}\hbox{$<$}}}\gamma
\mathrel{\hbox{\rlap{\hbox{\lower4pt\hbox{$\sim$}}}\hbox{$<$}}}
134^\circ$ arising from the usual fits of the unitarity 
triangle \cite{burasHF97}. If the ratio
\begin{equation}\label{Def-R}
R\equiv\frac{\mbox{BR}(B_d\to\pi^\mp K^\pm)}{\mbox{BR}(B^\pm\to\pi^\pm K)}
\end{equation}
is found to be smaller than 1 -- its present experimental range is 
$0.65\pm0.40$, so that this may indeed be the case -- the quantity 
$\gamma_0$ takes a maximal value 
\begin{equation}\label{gam-max}
\gamma_0^{\rm max}=\arccos(\sqrt{1-R})\,,
\end{equation} 
which depends only on $R$. In the future, these bounds may play an important
role to constrain the unitarity triangle (for a detailed study, see for 
instance \cite{gnps}). 

In addition to the general phase structure of the Standard Model of 
electroweak interactions, the following three assumptions have to be made
in order to arrive at (\ref{gam-max}):
\begin{itemize}
\item[i)]Isospin symmetry of strong interations can be used to derive 
relations between the QCD penguin amplitudes contributing to $B^+\to\pi^+ K^0$ 
and $B_d^0\to\pi^- K^+$.
\item[ii)]There is no non-trivial CP-violating weak phase present in the
$B^+\to\pi^+ K^0$ decay amplitude.
\item[iii)]Electroweak penguins play a negligible role in $B^+\to\pi^+ K^0$ 
and $B_d^0\to\pi^- K^+$. 
\end{itemize}
While the use of the $SU(2)$ isospin symmetry is certainly on solid theoretical
ground -- although special care has to be taken when applying it to 
penguin topologies with internal up quarks -- the other two assumptions 
are significantly less reliable and deserve further investigation.

As far as point (ii) is concerned, long-distance final-state interaction 
effects \cite{FSI}, which are related, for instance, to rescattering 
processes such as $B^+\to\{\pi^0K^+\}\to\pi^+K^0$, may in principle affect 
this assumption
about the $B^+\to\pi^+ K^0$ decay amplitude \cite{Bloketal}--\cite{atso}. An
implication of such effects may be sizeable CP violation as large as
${\cal O}(10\%)$ in $B^+\to\pi^+ K^0$ \cite{fknp}, while estimates based 
on simple quark-level calculations following the approach proposed by
Bander, Silverman and Soni \cite{bss} typically yield CP asymmetries of 
${\cal O}(1\%)$ \cite{pert-pens}. Reliable calculations of such rescattering 
effects are, however, very challenging and require theoretical insights into 
the dynamics of strong interactions, which unfortunately are not available at
present. In this paper we will therefore not make another attempt to
``calculate'' these effects. We rather include them in a completely 
general way in the formulae discussed in Sections~\ref{ampls} and 
\ref{gamma-det}, and advocate the use of experimental data on the decays
$B^+\to\pi^+ K^0$ and $B^+\to K^+\overline {K^0}$ to deal with final-state 
interactions. Employing $SU(3)$ flavour symmetry, these channels 
allow us to control the rescattering effects in the bounds on $\gamma$
completely. Interestingly, if there are indeed large contributions from such
rescattering processes, the combined $B^\pm\to K^\pm K$ branching ratio
may well be enhanced from its ``short-distance'' value ${\cal O}(10^{-6})$ 
to the $10^{-5}$ level, so that an experimental study of this mode appears
to be feasible at future $B$ factories. 

The role of electroweak penguins in non-leptonic $B$ decays, and strategies
to extract CKM phases, has been discussed extensively in the literature 
during the recent years \cite{rev}. In the decays $B^+\to\pi^+ K^0$ and 
$B_d^0\to\pi^- K^+$, electroweak penguins contribute only in 
``colour-suppressed'' form; model calculations using ``factorization'' 
to estimate the relevant hadronic matrix elements give contributions
at the 1\% level, and hence support point (iii) listed above \cite{fm2}. 
Such a treatment of the electroweak penguins may, however, underestimate 
their importance \cite{groro,neubert,fm3}, and it is therefore highly 
desirable to find more advanced methods to deal with these topologies. 
Similarly to our treatment of the rescattering effects, we also include the 
electroweak penguins in a completely general way in our formulae without 
making any approximation. As we will see below, in the case of the 
electroweak penguins it is, however, possible to improve their theoretical 
description considerably by using the relevant four-quark operators and the 
isospin symmetry of strong interactions to relate the corresponding hadronic 
matrix elements.  

The outline of this paper is as follows: in Section~\ref{ampls} we 
introduce a parametrization of the $B^+\to\pi^+ K^0$ and 
$B_d^0\to\pi^- K^+$ decay amplitudes in terms of ``physical'' quantities, 
relying only on the phase structure of the Standard Model and the isospin 
symmetry of strong interactions. In particular, the usual terminology 
of ``tree'' and QCD penguin amplitudes to describe these decays is clarified. 
In Section~\ref{gamma-det} we discuss strategies to constrain and determine 
the CKM angle $\gamma$ by using $B^\pm\to\pi^\pm K$ and $B_d\to\pi^\mp K^\pm$ 
decays. The bounds on $\gamma$ derived in \cite{fm2}, making use of only the 
combined branching ratios (\ref{BR-char}) and (\ref{BR-neut}), are generalized 
by taking into account in addition the CP-violating asymmetry arising in 
$B_d\to\pi^\mp K^\pm$, and transparent formulae including both final-state 
interactions and electroweak penguins in a completely general way are 
presented. In Section~\ref{FSI-effects} we investigate the role of 
rescattering processes in these strategies and point out that the modes 
$B^\pm\to\pi^\pm K$ and $B^\pm\to K^\pm K$ allow us to include these effects 
in the bounds on $\gamma$ completely by using the $SU(3)$ flavour symmetry. A 
detailed analysis of the electroweak penguin effects is performed in 
Section~\ref{EWP-effects}, where also an improved theoretical treatment of 
the corresponding contributions and a first step to constrain them 
experimentally with the help of the decay $B^+\to\pi^+\pi^0$ are presented. 
The combined effects of final-state interactions and electroweak penguins are 
discussed in Section~\ref{comb-eff} by considering a few selected examples, 
and the conclusions are summarized in Section~\ref{conclu}.

\boldmath
\section{General Description of $B^\pm\to\pi^\pm K$ and 
$B_d\to\pi^\mp K^\pm$ within the Standard Model}\label{ampls}
\unboldmath
The subject of this section is a general description of the decays 
$B^\pm\to\pi^\pm K$ and $B_d\to\pi^\mp K^\pm$ within the framework of
the Standard Model. After a parametrization of their decay amplitudes, 
expressions for the observables provided by these modes are given,
taking into account both rescattering and electroweak penguin effects.

\boldmath
\subsection{The $B^+\to\pi^+K^0$ and $B^0_d\to\pi^-K^+$ Decay 
Amplitudes}\label{dec-ampls}
\unboldmath
Let us have a closer look at the charged $B$ decay $B^+\to\pi^+ K^0$ first. 
Its transition amplitude can be written as
\begin{equation}
A(B^+\to\pi^+K^0)=\lambda^{(s)}_u(P_u+P_{\rm ew}^u+{\cal A})+
\lambda^{(s)}_c(P_c+P_{\rm ew}^c)+\lambda^{(s)}_t(P_t+P_{\rm ew}^t)\,,
\end{equation}
where $P_q$ and $P_{\rm ew}^q$ denote contributions from QCD and
electroweak penguin topologies with internal $q$ quarks $(q\in\{u,c,t\})$, 
respectively; ${\cal A}$ is related to annihilation topologies, and
\begin{equation} 
\lambda^{(s)}_q\equiv V_{qs}V_{qb}^\ast
\end{equation}
are the usual CKM factors. Making use of the unitarity of the CKM matrix 
and applying the Wolfenstein parametrization \cite{wolf} yields
\begin{equation}\label{Bpampl}
A(B^+\to\pi^+K^0)=-\left(1-\frac{\lambda^2}{2}\right)\lambda^2A\left[
1+\rho\,e^{i\theta}e^{i\gamma}\right]{\cal P}_{tc}\,,
\end{equation}
where
\begin{equation}
{\cal P}_{tc}\equiv\left|{\cal P}_{tc}\right|e^{i\delta{tc}}=
\left(P_t-P_c\right)+(P_{\rm ew}^t-P_{\rm ew}^c)
\end{equation}
and
\begin{equation}\label{rho-def}
\rho\,e^{i\theta}=\frac{\lambda^2R_b}{1-\lambda^2/2}
\left[1-\left(\frac{{\cal P}_{uc}+{\cal A}}{{\cal P}_{tc}}\right)\right]
\end{equation}
with
\begin{equation}
{\cal P}_{uc}=\left(P_u-P_c\right)+\left(P_{\rm ew}^u-P_{\rm ew}^c\right)\,.
\end{equation}
In these expressions, $\delta_{tc}$ and $\theta$ denote CP-conserving strong
phases, and the present status of the relevant CKM factors is given by
\begin{equation}
\lambda\equiv|V_{us}|=0.22\,,\quad
A\equiv\frac{1}{\lambda^2}\left|V_{cb}\right|=0.81\pm0.06\,,\quad
R_b\equiv\frac{1}{\lambda}\left|\frac{V_{ub}}{V_{cb}}\right|=0.36\pm0.08\,.
\end{equation}

On the other hand, the decay amplitude of $B^0_d\to\pi^-K^+$ takes the form
\begin{equation}
A(B^0_d\to\pi^-K^+)=-\left[\lambda^{(s)}_u(\tilde P_u+\tilde P_{\rm ew}^u+
\tilde{\cal T}\,)+\lambda^{(s)}_c(\tilde P_c+\tilde P_{\rm ew}^c)+
\lambda^{(s)}_t(\tilde P_t+\tilde P_{\rm ew}^t)\right],
\end{equation}
where the notation is as in (\ref{Bpampl}) and the minus sign is due to our 
definition of meson states. The amplitude $\tilde{\cal T}$ arises from the 
fact that the current--current operators 
\begin{equation}\label{CC-ops}
Q_1^u=(\bar u_{\alpha} s_{\beta})_{{\rm V-A}}\;(\bar b_{\beta} 
u_{\alpha})_{{\rm V-A}}\,,\quad
Q_2^u=(\bar u_{\alpha} s_{\alpha})_{{\rm V-A}}\;
(\bar b_{\beta} u_{\beta})_{{\rm V-A}}\,, 
\end{equation}
where $\alpha$ and $\beta$ are colour indices, contribute to 
$B^0_d\to\pi^-K^+$ also through insertions into tree-diagram-like 
topologies. Such contributions are absent in the case of $B^+\to\pi^+K^0$,
where these operators contribute only through insertions into penguin and
annihilation topologies, which are described by $P_u$ and ${\cal A}$,
respectively \cite{bfm}.

\vspace{0.3truecm}

Using the $SU(2)$ isospin symmetry, which implies the relations
\begin{equation}
\tilde P_c=P_c\,,\quad \tilde P_t=P_t
\end{equation}
for the penguin topologies with internal charm and top quarks \cite{bfm}, 
we arrive at the following amplitude relations:
\begin{eqnarray}
A(B^+\to\pi^+K^0)&\equiv&P\label{ampl-neut}\\
A(B^0_d\to\pi^-K^+)&=&-\,[P+T+P_{\rm ew}]\label{ampl-char}\,,
\end{eqnarray}
where
\begin{equation}\label{T-def}
T\equiv|T|\,e^{i\delta_T}e^{i\gamma}=\lambda^4A\,R_b\left[
\tilde{\cal T}-{\cal A}+\left(\tilde P_u-P_u\right)+\left(\tilde
P_{\rm ew}^u-\tilde P_{\rm ew}^t\right)-\left(P_{\rm ew}^u-
P_{\rm ew}^t\right)\right]e^{i\gamma}
\end{equation}
\begin{equation}\label{Pew-def}
P_{\rm ew}\equiv-\,|P_{\rm ew}|\,e^{i\delta_{\rm ew}}=-\left(1-
\frac{\lambda^2}{2}\right)\lambda^2A\left[\left(\tilde P_{\rm ew}^t-\tilde
P_{\rm ew}^c\right)-\left(P_{\rm ew}^t-P_{\rm ew}^c\right)\right].
\end{equation}
Here $\delta_T$ and $\delta_{\rm ew}$ denote CP-conserving strong phases. 
The amplitude relations (\ref{ampl-neut}) and (\ref{ampl-char}), 
which play a central role to obtain information about the CKM angle 
$\gamma$, rely only on the isospin symmetry of strong interactions (for 
a detailed discussion, see \cite{bfm}) and involve only ``physical'', 
i.e.\ renormalization-scale- and scheme-independent, quantities. This feature 
is obvious since $P$, usually referred to as a $\bar b\to\bar s$ penguin
amplitude, is {\it defined} through the $B^+\to\pi^+K^0$ decay amplitude
given in (\ref{Bpampl}). The combination $-\,(T+P_{\rm ew})$ is on the other
hand {\it defined} through the sum of the $B^+\to\pi^+K^0$ and 
$B^0_d\to\pi^-K^+$ decay amplitudes and therefore also a ``physical'' 
quantity. Since $T$ and $P_{\rm ew}$ describe two different CKM 
contributions related to the weak phase factors 
$e^{i\gamma}$ and $e^{i\pi}=-1$, respectively, they are ``physical'' 
amplitudes as well. A similar comment applies to the quantities 
$\rho\,e^{i\theta}$ and ${\cal P}_{tc}$ parametrizing the 
$B^+\to\pi^+K^0$ decay amplitude in (\ref{Bpampl}). Let us note that $T$ 
is usually referred to as a ``tree'' amplitude. As can be seen in 
(\ref{T-def}), $T$ actually receives not only such ``tree'' contributions 
corresponding to $\tilde{\cal T}$, but also contributions from annihilation 
and penguin topologies.

\boldmath
\subsection{The $B^\pm\to\pi^\pm K$ and $B_d\to\pi^\mp K^\pm$
Observables}\label{observables}
\unboldmath
Taking into account that the amplitude relations for the CP-conjugate modes
can be obtained straightforwardly from the expressions given in 
Subsection~\ref{dec-ampls} by performing the substitution 
$\gamma\to-\,\gamma$, and introducing the observables
\begin{equation}\label{r-eps-def}
r\equiv\frac{|T|}{\sqrt{\left\langle|P|^2\right\rangle}}\,,\quad
\epsilon\equiv\frac{|P_{\rm ew}|}{\sqrt{\left\langle|P|^2\right\rangle}}\,,
\end{equation}
where 
\begin{equation}
\left\langle|P|^2\right\rangle\equiv\frac{1}{2}\left(|P|^2+|\overline{P}|^2
\right),
\end{equation}
as well as the CP-conserving strong phases
\begin{equation}
\delta\equiv\delta_T-\delta_{tc}\,,
\quad\Delta\equiv\delta_{\rm ew}-\delta_{tc}\,,
\end{equation}
we get the following expression for the ratio $R$ of combined branching
ratios, which has been defined in~(\ref{Def-R}):
\begin{eqnarray}
\lefteqn{R=1-2\,r\,\frac{\left[\,\cos\delta\cos\gamma+\rho\,\cos(\delta-\theta)
\,\right]}{\sqrt{1+2\,\rho\,\cos\theta\cos\gamma+\rho^2}}+r^2}\nonumber\\
&&+\,2\,\epsilon\,\frac{\left[\,\cos\Delta+\rho\,\cos(\Delta-\theta)\cos\gamma
\,\right]}{\sqrt{1+2\,\rho\,\cos\theta\cos\gamma+\rho^2}}-2\,r\,\epsilon\,
\cos(\delta-\Delta)\cos\gamma+\epsilon^2.\label{R-exp}
\end{eqnarray}
In order to determine $\gamma$, the ``pseudo-asymmetry''
\begin{equation}
A_0\equiv\frac{\mbox{BR}(B^0_d\to\pi^-K^+)-\mbox{BR}(\overline{B^0_d}\to
\pi^+K^-)}{\mbox{BR}(B^+\to\pi^+K^0)+\mbox{BR}(B^-\to\pi^-\overline{K^0})}
\end{equation}
turns out to be very useful \cite{groro}. It takes the form
\begin{equation}\label{A0-exp}
A_0=A_++\frac{2\,r\,\sin\delta\sin\gamma}{\sqrt{1+
2\,\rho\,\cos\theta\cos\gamma+\rho^2}}+2\,r\,\epsilon\sin(\delta-\Delta)
\sin\gamma+\frac{2\,\epsilon\,\rho\,\sin(\Delta-\theta)\sin\gamma}{\sqrt{1+
2\,\rho\,\cos\theta\cos\gamma+\rho^2}},
\end{equation}
where
\begin{equation}\label{Ap-def}
A_+\equiv\frac{\mbox{BR}(B^+\to\pi^+K^0)-\mbox{BR}(B^-\to\pi^-
\overline{K^0})}{\mbox{BR}(B^+\to\pi^+K^0)+\mbox{BR}(B^-\to\pi^-
\overline{K^0})}=-\,\frac{2\,\rho\,\sin\theta\sin\gamma}{1+
2\,\rho\,\cos\theta\cos\gamma+\rho^2}
\end{equation}
measures direct CP violation in the decay $B^+\to\pi^+K^0$. Note that tiny 
phase-space effects have been neglected in (\ref{R-exp}) and (\ref{A0-exp})
(for a more detailed discussion, see \cite{fm2}). 

The expressions given in (\ref{r-eps-def})--(\ref{R-exp}) are the
correct generalization of the formulae derived in \cite{fm2}, where points
(ii) and (iii) listed in Section~\ref{intro} have been assumed, i.e.\
$\rho=0$ and $\epsilon=0$. They take into account both rescattering and
electroweak penguin effects in a completely general way and make use only
of the isospin symmetry of strong interactions. Before we investigate
these effects in detail in Sections~\ref{FSI-effects} and \ref{EWP-effects},
let us first discuss strategies to constrain and determine the CKM angle 
$\gamma$ from these observables.

\boldmath
\section{Strategies to Constrain and Determine the CKM Angle $\gamma$ from 
$B^\pm\to\pi^\pm K$ and $B_d\to\pi^\mp K^\pm$ Decays}\label{gamma-det}
\unboldmath
The observables $R$ and $A_0$ provide valuable information about the 
CKM angle $\gamma$. A measurement of the asymmetry $A_0$ allows us to 
eliminate the CP-conserving strong phase $\delta$ in the ratio $R$ of 
combined $B\to\pi K$ branching ratios (see (\ref{R-exp}) and (\ref{A0-exp})).
In the special case $\rho=\epsilon=0$, we get an expression for $R$ 
depending only on $\gamma$ and $r$. Consequently, if $r$
could be fixed, we would have a method to determine $\gamma$ 
\cite{PAPIII,groro}. While this approach was presented in \cite{PAPIII} 
as an approximate way to fix this angle, because of the model dependence 
introduced through $r$, recent studies \cite{groro,wuegai} using arguments 
based on ``factorization''  came to the conclusion that a future theoretical 
uncertainty of $r$ as small as ${\cal O}(10\%)$ may be achievable. In that 
case, a determination of $\gamma$ at future $B$ factories (BaBar, BELLE, 
CLEO III) employing this approach would be limited rather by statistics than 
by the uncertainty introduced through $r$ \cite{wuegai}. In 
Section~\ref{FSI-effects} we will see that large contributions to $B\to\pi K$
decays from rescattering processes may affect $r$ severely, thereby shifting 
it significantly from its ``factorized'' value \cite{fm2}
\begin{equation}\label{r-fact}
\left.r\right|_{\rm fact}=0.16\times a_1\times
\left[\frac{|V_{ub}|}{3.2\times10^{-3}}
\right]\times\sqrt{\left[\frac{2.3\times10^{-5}}{\mbox{BR}(B^\pm 
\to \pi^\pm K)}\right]\times\left[\frac{\tau_{B_u}}{1.6\,\mbox{ps}}\right]}\,.
\end{equation}
Here the relevant $B\to\pi$ form factor obtained in the BSW model 
\cite{BSW} has been used and $a_1\approx1$ is the usual phenomenological 
colour factor \cite{ns}. Consequently a reliable determination of $r$
may be precluded by rescattering effects.

\boldmath
\subsection{Bounds on the CKM Angle $\gamma$}\label{gam-bounds}
\unboldmath
As was pointed out in \cite{fm2}, the observable $R$ by itself may allow
an interesting bound on $\gamma$, which does {\it not} require any information 
about $r$. The idea is as follows: if one assumes $\rho=\epsilon=0$ and
keeps both $r$ and the strong phase $\delta$ in the expression for $R$
as free ``unknown'' parameters, one finds that it takes a minimal value given 
by $\sin^2\gamma$, i.e.\ we have $R\ge\sin^2\gamma$. Consequently, if $R$ is
found experimentally to be smaller than 1, we get the allowed range
(\ref{gamma-bound1}) for $\gamma$, with $\gamma_0$ given by (\ref{gam-max}).

\begin{figure}
\centerline{
\rotate[r]{
\epsfxsize=9.2truecm
\epsffile{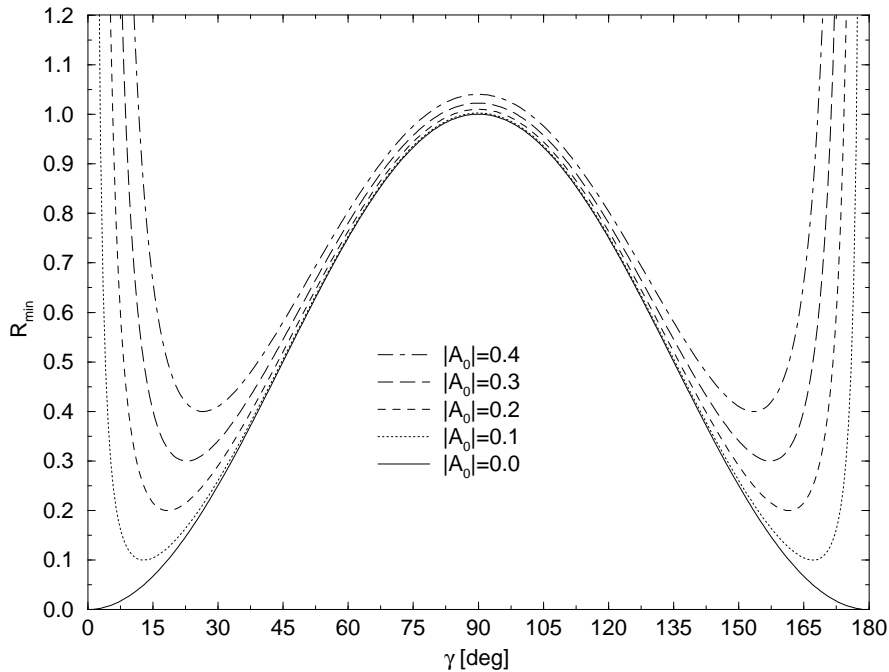}}}
\caption{The dependence of $R_{\rm min}$ on the CKM angle $\gamma$ for 
various values of $A_0$ in the case of neglected rescattering and electroweak 
penguin effects, i.e.\ $\rho=\epsilon=0$.}\label{fig:Rmin}
\end{figure}

In this section we improve this bound in two respects by taking into
account both rescattering and electroweak penguin effects, and the asymmetry
$A_0$. As we have already noted, this observable allows us to eliminate the
strong phase  $\delta$ in (\ref{R-exp}). To this end, we rewrite 
(\ref{R-exp}) and (\ref{A0-exp}) as
\begin{equation}\label{R-simpl}
R=R_0-2\,r\,(h\cos\delta+k\sin\delta)+r^2
\end{equation}
and
\begin{equation}\label{A-Def}
A=\left(B\sin\delta-C\cos\delta\right)r\,,
\end{equation}
respectively, where the quantities
\begin{equation}
R_0=1+2\,\frac{\epsilon}{w}\,\left[\,\cos\Delta+\rho\,\cos(\Delta-\theta)
\cos\gamma\,\right]+\epsilon^2
\end{equation}
\begin{equation}
h=\frac{1}{w}\left(\cos\gamma+\rho\,\cos\theta\right)+\epsilon\,\cos\Delta
\,\cos\gamma\,,
\quad k=\frac{\rho}{w}\,\sin\theta+\epsilon\,\sin\Delta\,\cos\gamma\,,
\end{equation}
with 
\begin{equation}
w=\sqrt{1+2\,\rho\,\cos\theta\cos\gamma+\rho^2},
\end{equation}
and
\begin{equation}
A=\frac{A_0-A_+}{2\,\sin\gamma}-\frac{\epsilon\,\rho}{w}
\sin(\Delta-\theta)\,,\quad
B=\frac{1}{w}+\epsilon\,\cos\Delta\,,\quad
C=\epsilon\,\sin\Delta\,,
\end{equation}
are independent of $r$. A straightforward calculation yields the expressions
\begin{equation}
\sin\delta=\frac{AB\pm C\,\sqrt{(B^2+C^2)\,r^2-A^2}}{(B^2+C^2)\,r}\,,\quad
\cos\delta=\frac{-AC\pm B\,\sqrt{(B^2+C^2)\,r^2-A^2}}{(B^2+C^2)\,r}\,,
\end{equation}
allowing the elimination of the strong phase $\delta$ in (\ref{R-simpl}):
\begin{equation}\label{R-elim}
R=R_0-AD\mp E\sqrt{(B^2+C^2)\,r^2-A^2}+r^2,
\end{equation}
where 
\begin{equation}
D=2\left(\frac{k\,B-h\,C}{B^2+C^2}\right),\quad
E=2\left(\frac{h\,B+k\,C}{B^2+C^2}\right).
\end{equation}
Treating now $r$ in (\ref{R-elim}) as a free variable, we find that $R$ 
takes a minimal value for
\begin{equation}
r=r_0\equiv\sqrt{\frac{A^2}{B^2+C^2}\,+\,\frac{(B^2+C^2)E^2}{4}}\,,
\end{equation}
which has the following form:
\begin{equation}\label{Rmin}
R_{\rm min}=\kappa\,\sin^2\gamma\,+\,
\frac{1}{\kappa}\left(\frac{A_0}{2\,\sin\gamma}\right)^2.
\end{equation}
The rescattering and electroweak penguin effects are included in this 
transparent expression through the parameter $\kappa$, which is given by
\begin{equation}
\kappa=\frac{1}{w^2}\left[\,1+2\,(\epsilon\,w)\cos\Delta+
(\epsilon\,w)^2\,\right].
\end{equation}
In order to derive these formulae, no approximations have been made and they
are valid exactly. The dependence of $R_{\rm min}$ on $\gamma$ for the
special case $\rho=\epsilon=0$ and for various values of $A_0$ is shown in 
Fig.~\ref{fig:Rmin}, where $A_0=0$ corresponds to the bound presented 
in \cite{fm2}. The modifications of this figure through $\rho\not=0$ and 
$\epsilon\not=0$ are investigated in the following sections. If $R$ will
be found experimentally to be smaller than 1, or if it should become possible 
to obtain an experimental upper limit $R_{\rm exp}^{\rm max}<1$, the 
values of $\gamma$ implying $R_{\rm min}>R$ or 
$R_{\rm min}>R_{\rm exp}^{\rm max}$ would be excluded. For values of $R$
as small as 0.65, which is the central value of present CLEO data, a large 
region around $\gamma=90^\circ$ would be excluded. As soon as we have a
non-vanishing experimental result for $A_0$, also an interval around 
$\gamma=0^\circ$ and $180^\circ$ can be ruled out, while the impact on the 
excluded region around $90^\circ$ is rather small, as can be seen in 
Fig.~\ref{fig:Rmin}. Let us note that the minima of the curves shown in this 
figure correspond to $R_{\rm min}=|A_0|$, and that the values of 
$\gamma$ between $0^\circ$ and $90^\circ$ correspond to $\cos\delta>0$, 
while those between $90^\circ$ and $180^\circ$ to $\cos\delta<0$. Estimates 
based on quark-level calculations indicate $\cos\delta>0$ and hence
favour the former range \cite{fm2,ag}.

\boldmath
\subsection{Bounds on $r$ and the Determination of the CKM 
Angle $\gamma$}\label{r-bounds}
\begin{figure}
\centerline{
\rotate[r]{
\epsfxsize=9.2truecm
\epsffile{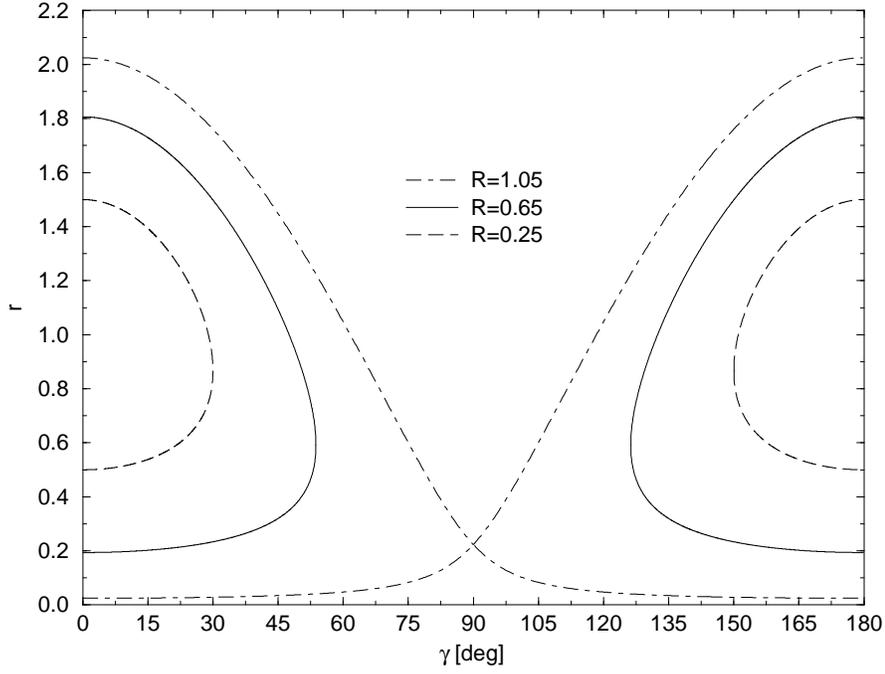}}}
\caption{The allowed regions for $r$ corresponding to various values 
of $R$ in the case of neglected rescattering and electroweak penguin 
effects.}\label{fig:r-bound}
\end{figure}

\unboldmath
Another interesting aspect of a measurement of $R$ is a bound on $r$ of
the form
\begin{equation}
r_{\rm min}\le r\le r^{\rm max}
\end{equation}
with
\begin{equation}
r_{\rm min}^{\rm max}=\left|\sqrt{R_0\,-\,\kappa\,\sin^2\gamma}\,
\pm\,\sqrt{R\,-\,\kappa\,\sin^2\gamma}\right|,
\end{equation}
which arises if we treat $\delta$ in (\ref{R-simpl}) as a free 
parameter. This bound is shown in Fig.~\ref{fig:r-bound} for 
$\rho=\epsilon=0$ and for various values of $R$ corresponding to its presently 
allowed experimental range. In this figure it can also be seen nicely 
which values of $\gamma$ are excluded in the case of $R<1$. Moreover, we 
observe that small values of $R$ require large values of $r$ in comparison
with the ``factorized'' result (\ref{r-fact}), which is at the edge of 
compatibility with the central values of the present CLEO measurements 
\cite{cleo}, yielding $R=0.65$ and $r\ge0.2$. For $\gamma
\mathrel{\hbox{\rlap{\hbox{\lower4pt\hbox{$\sim$}}}\hbox{$>$}}}41^\circ$ --
the lower bound obtained from the usual fits of the unitarity triangle
\cite{burasHF97} -- we even have $r
\mathrel{\hbox{\rlap{\hbox{\lower4pt\hbox{$\sim$}}}\hbox{$>$}}}0.3$.
This interesting feature has already been pointed out in \cite{fm2}, and we 
shall come back to it in the following section.

\begin{figure}
\centerline{
\rotate[r]{
\epsfxsize=9.2truecm
\epsffile{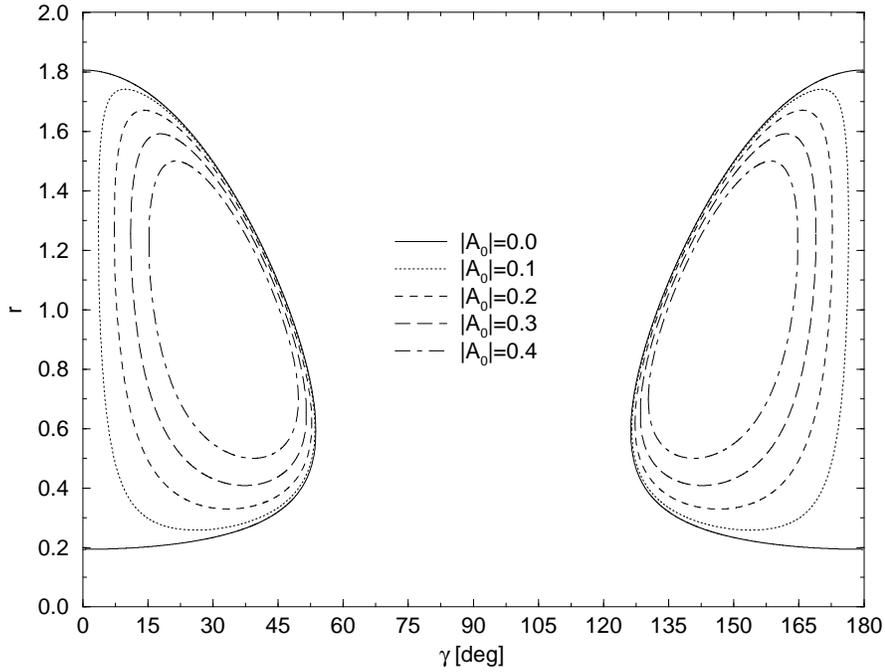}}}
\caption{The dependence of $r$ on the CKM angle $\gamma$ for $R=0.65$ 
and for various values of the asymmetry $A_0$ in the case of neglected 
rescattering and electroweak penguin effects.}\label{fig:r-det065}
\end{figure}

As soon as $A_0$ has been measured, we can go beyond this bound. Then we 
are in a position to determine the dependence of $r$ on $\gamma$ 
with the help of (\ref{R-elim}), yielding
\begin{equation}
r=\sqrt{r_0^2\,+\,\left(R-R_{\rm min}\right)\,\pm\,\sqrt{(B^2+C^2)\,
E^2\left(R-R_{\rm min}\right)}}\,.
\end{equation}
In Fig.~\ref{fig:r-det065} we have chosen $R=0.65$ to illustrate this 
dependence for various pseudo-asymmetries $A_0$ in the case of neglected
rescattering and electroweak penguin effects. The curves plotted there are 
a mathematical implementation of the amplitude triangles proposed in 
\cite{PAPIII}. Once $R$ and $A_0$ have been measured, the corresponding 
contours in the $\gamma$--$r$ plane can be calculated. If $r$ could be 
fixed by using an additional input, $\gamma$ could be determined up to a 
four-fold ambiguity, as can be seen in Fig.~\ref{fig:r-det065}. For $R<1$, 
also the allowed range for $\gamma$ can be read off nicely from the 
corresponding contours. 

Although the formulae derived in this section are completely general, taking
into account both rescattering and electroweak penguin effects, we have not
included these contributions in Figs.~\ref{fig:Rmin}--\ref{fig:r-det065} in
order to illustrate the strategies to constrain and determine the CKM angle 
$\gamma$ from $B^\pm\to\pi^\pm K$, $B_d\to\pi^\mp K^\pm$ decays in a 
transparent way. The issue of final-state interactions and electroweak 
penguins in these methods will be the subject of the remainder of this
paper.

\boldmath
\section{The Role of Rescattering Processes}\label{FSI-effects}
\unboldmath
In the formulae derived in the previous two sections, contributions from
rescattering processes are included through the quantity $\rho\,e^{i\theta}$
introduced in (\ref{rho-def}). An important implication of $\rho\not=0$ and
$\theta\not\in\{0,\pi\}$ is direct CP violation in the mode $B^+\to\pi^+K^0$,
as can be seen in (\ref{Ap-def}). The parameter $\rho$ describing the
``strength'' of the rescattering effects is, however, highly CKM-suppressed
by $\lambda^2R_b\approx0.02$. In (\ref{rho-def}), we have to distinguish 
between contributions from penguin topologies with internal top, charm and 
up quarks, and annihilation topologies. Concerning the hierarchy of these 
contributions, the usual expectation is that annihilation processes play a 
very minor role and that penguins with internal top quarks are the most 
important ones. Following these considerations, one would expect 
$\rho\approx\lambda^2R_b$. However, also penguins with internal charm and up 
quarks lead in general to important contributions, which cannot be neglected 
\cite{fbf,ital}. Model calculations performed at the perturbative quark-level 
to estimate these contributions give $\rho={\cal O}(1\%)$ and do not indicate
a significant compensation of the very large CKM suppression of 
$\rho$.

\boldmath
\subsection{A Closer Look at the Rescattering Effects}\label{FSI-gen}
\unboldmath
It has recently been discussed in \cite{bfm} that rescattering processes of 
the kind
\begin{equation}\label{c-res}
B^+\to\{F_c^{(s)}\}\to\pi^+K^0,
\end{equation}
where $F_c^{(s)}\in\{\overline{D^0}D_s^+,\,\overline{D^0}D_s^{\ast+},\,
\overline{D^{\ast 0}}D_s^{\ast+},\,\ldots\}$, are related to penguin 
topologies with internal charm quarks, while rescattering processes of the
kind 
\begin{equation}\label{u-res}
B^+\to\{F_u^{(s)}\}\to\pi^+K^0,
\end{equation}
where $F_u^{(s)}\in\{\pi^0K^+,\,\pi^0K^{\ast +},\,\rho^0K^{\ast +},\,
\ldots\}$, are related to penguin topologies with internal up quarks and
to annihilation topologies, which will be discussed below. These
final-state-interaction effects, where channels originating from the 
current--current operators $Q_{1,2}^c$ and $Q_{1,2}^u$ through insertions into
tree-diagram-like topologies are involved, can be considered as long-distance 
contributions to the amplitudes $P_c$ and $P_u$, respectively, and are included
this way in (\ref{rho-def}). While we would have $\rho\approx0$ if 
rescattering processes of type (\ref{c-res}) played the dominant role 
in $B^+\to\pi^+ K^0$, or $\rho={\cal O}(\lambda^2R_b)$ if both (\ref{c-res}) 
and (\ref{u-res}) were similarly important, $\rho$ would be as large as 
${\cal O}(10\%)$ if the final-state interactions arising from processes such 
as (\ref{u-res}) would dominate $B^+\to\pi^+ K^0$ so that $|{\cal P}_{uc}|/
|{\cal P}_{tc}|={\cal O}(5)$. This order of magnitude is found in a recent 
attempt \cite{fknp} to evaluate the rescattering  processes (\ref{u-res})
using Regge phenomenology. 

The usual argument for the suppression of annihilation processes relative 
to tree-diagram-like topologies by a factor $f_B/m_B$ does not apply to 
rescattering processes \cite{Bloketal,neubert}. Consequently, these 
topologies may also play a more important role than na\"\i vely expected. 
Model calculations \cite{Bloketal} based on Regge phenomenology typically 
give an enhancement of the ratio $|{\cal A}|/|\tilde{\cal T}|$ from 
$f_B/m_B\approx0.04$ to ${\cal O}(0.2)$. Rescattering processes of this 
kind can be probed, e.g.\ by the $\Delta S$\,=\,0 decay $B^0_d\to K^+K^-$. 
A future stringent bound on BR$(B^0_d\to K^+K^-)$ at the level of 
${\cal O}(10^{-7})$ or lower may provide a useful limit on these rescattering 
effects~\cite{groro}. The present upper bound obtained by the CLEO 
collaboration is $4.3\times10^{-6}$ \cite{cleo}.

Although the ``factorization'' hypothesis \cite{facto} is in general 
questionable, it may work reasonably well for the colour-allowed amplitude 
$\tilde{\cal T}$ \cite{bjorken}. Consequently, in contrast to $r$ defined 
by (\ref{r-eps-def}), the quantity
\begin{equation}\label{Def-r-tilde}
\tilde r\equiv\lambda^4A\,R_b\,\frac{|\tilde{\cal T}|}{\sqrt{\left
\langle|P|^2\right\rangle}}
\end{equation}
may be described rather well by the ``factorized'' expression (\ref{r-fact}), 
i.e.\ $\tilde r\approx0.15$. Since the intrinsic ``strength'' of decays such 
as $B^+\to\pi^0K^+$ representing the ``first step'' of the rescattering 
processes (\ref{u-res}) is given by  $\tilde r$, we have a ``plausible'' 
upper bound for $\rho$ through
$\rho\mathrel{\hbox{\rlap{\hbox{\lower4pt\hbox{$\sim$}}}\hbox{$<$}}}\tilde r
\approx0.15$. Note that $\rho$ is typically one order of magnitude smaller, 
i.e.\ $\rho={\cal O}(0.02)$, if rescattering processes do not play the 
dominant role in $B^+\to\pi^+ K^0$. 

In the following discussion we will not comment further on quantitative
estimates of rescattering effects. A reliable theoretical treatment is very 
difficult and requires insights into the dynamics of strong interactions 
that are unfortunately not available at present. In this paper we rather
investigate the sensitivity of the bounds on $\gamma$ presented in 
Section~\ref{gamma-det} on the quantity $\rho\,e^{i\theta}$, which 
parametrizes the rescattering processes, and advocate the use of 
experimental data to obtain insights into these final-state interactions. 

\boldmath
\subsection{Rescattering Effects in Bounds on $\gamma$}\label{FSI-gamma}
\begin{figure}
\centerline{
\rotate[r]{
\epsfxsize=9.2truecm
\epsffile{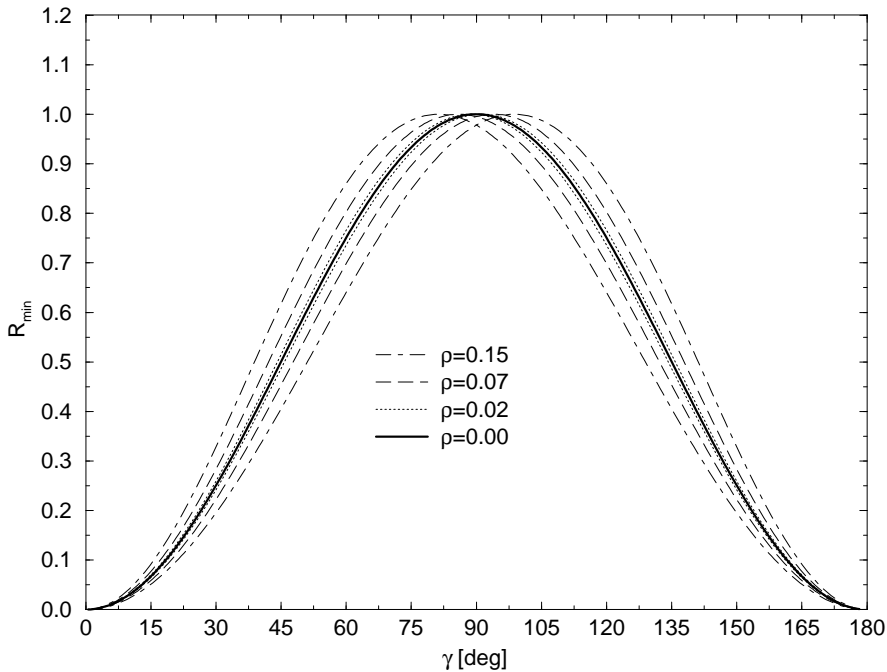}}}
\caption{The effect of final-state interactions on $R_{\rm min}$ for $A_0=0$.
The curves for a given value of $\rho$ correspond to 
$\theta\in\{0^\circ,180^\circ\}$ and represent the maximal shift from 
$\rho=0$.}\label{fig:Rminres0}
\end{figure}

\unboldmath
Considering only rescattering processes and neglecting electroweak penguin
contributions, which will be discussed in Section~\ref{EWP-effects}, 
Eq.\ (\ref{Rmin}) gives the simple expression
\begin{equation}\label{Rmin-simplFSI}
R_{\rm min}=\left(\frac{\sin\gamma}{w}\right)^2+\,\left(\frac{w\,A_0}{2\sin
\gamma}\right)^2,
\end{equation}
where the rescattering effects are included through 
$w=\sqrt{1+2\,\rho\,\cos\theta\cos\gamma+\rho^2}$. While these effects are 
minimal for $\theta\in\{90^\circ,270^\circ\}$ and only of second order, i.e.\ 
of ${\cal O}(\rho^2)$, they are maximal for $\theta\in\{0^\circ,180^\circ\}$.
In Fig.~\ref{fig:Rminres0} we show these maximal effects for various values 
of $\rho$ in the case of $A_0=0$. Looking at this figure, we first observe 
that we have negligibly small effects for $\rho=0.02$, which has been assumed
in \cite{fm2} in the form of point~(ii) listed in Section~\ref{intro}. For 
values of $\rho$ as large as $0.15$, we have an uncertainty for 
$\gamma_0^{\rm max}$ (see (\ref{gamma-bound1}) and (\ref{gam-max})) of at 
most $\pm\,10^\circ$. Consequently, even for large rescattering effects, 
a significant region around $\gamma=90^\circ$ will still be excluded, provided
$R$ is found experimentally to be smaller than 1, preferably close to its 
present central value of 0.65 or even smaller.

\begin{figure}
\centerline{
\rotate[r]{
\epsfxsize=9.2truecm
\epsffile{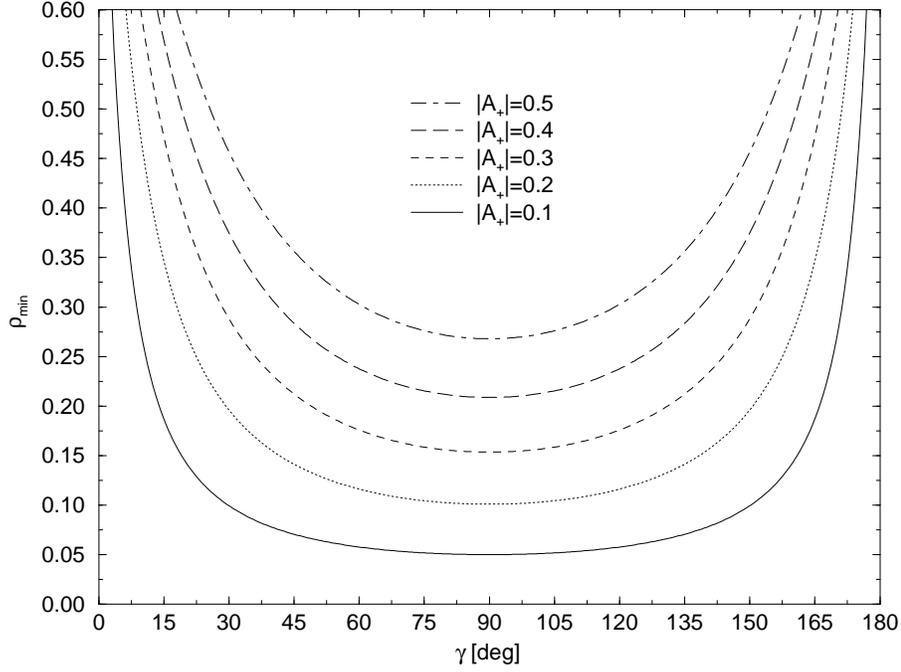}}}
\caption{The dependence of $\rho_{\rm min}$ on the CKM angle $\gamma$ 
for various values of the direct CP asymmetry $A_+$ arising in the decay
$B^+\to\pi^+K^0$.}\label{fig:rhomin}
\end{figure}

Since we have assumed $\theta\in\{0^\circ,180^\circ\}$ in 
Fig.~\ref{fig:Rminres0} to illustrate the maximal effect on $R_{\rm min}$
arising from rescattering processes described by a given value of $\rho$, 
the decay $B^+\to\pi^+K^0$ would exhibit no direct CP violation in this case. 
However, as soon as a non-vanishing value of $A_+$ has been measured, we 
are in a position to eliminate the CP-conserving strong phase $\theta$ in 
$R_{\rm min}$. Introducing
\begin{equation}
U=\frac{A_+^2W\cos\gamma}{\sin^2\gamma+A_+^2\cos^2\gamma}\,,\quad
V=\frac{A_+^2W^2-\sin^2\gamma}{\sin^2\gamma+A_+^2\cos^2\gamma}\,,
\end{equation}
with
\begin{equation}
W=\frac{1+\rho^2}{2\,\rho}\,,
\end{equation}
and using (\ref{Ap-def}), we obtain
\begin{equation}\label{Theta-deter}
\cos\theta=-\,U\pm\,\sqrt{U^2-V},
\end{equation}
which allows us to fix $w$ and 
\begin{equation}
\sin\theta=-\,\frac{w^2\,A_+}{2\,\rho\,\sin\gamma}
\end{equation}
up to a two-fold ambiguity. Moreover an interesting constraint on $\rho$ 
is provided by the direct CP asymmetry $A_+$ in $B^+\to\pi^+K^0$. It implies 
an allowed range 
\begin{equation}\label{rho-bound}
\rho_{\rm min}\,\leq\,\rho\,\leq\,\rho^{\rm max},
\end{equation}
where the upper and lower bounds
\begin{equation}\label{rho-min-max}
\rho_{\rm min}^{\rm max}=\frac{\sqrt{A_+^2\,+\,
\left(1-A_+^2\right)\sin^2\gamma}\,\pm\,\sqrt{\left(1-A_+^2\right)
\sin^2\gamma}}{|A_+|}
\end{equation}
correspond to $\theta=\theta_0$ with
\begin{equation}
\cos\theta_0=-\,\frac{|A_+|\cos\gamma}{\sqrt{\sin^2\gamma+
A_+^2\cos^2\gamma}}\,,\quad
\sin\theta_0=\frac{\tan\gamma}{A_+}\,\cos\theta_0\,.
\end{equation}
Keeping the CKM angle $\gamma$ as a free parameter, we find 
\begin{equation}
\rho\,\geq\,\frac{1-\sqrt{1-A_+^2}}{|A_+|}\,=\,
\left.\rho_{\rm min}\right|_{\gamma=90^\circ}.
\end{equation}
In particular the lower bound $\rho_{\rm min}$ is of special interest, and we 
show its dependence on the CKM angle $\gamma$ for various values of $A_+$ in 
Fig.~\ref{fig:rhomin}. It is interesting to note that these curves exclude
values of $\gamma$ around $0^\circ$ and $180^\circ$, if an upper limit on
$\rho$ is available. 

\begin{figure}
\centerline{
\rotate[r]{
\epsfxsize=9.2truecm
\epsffile{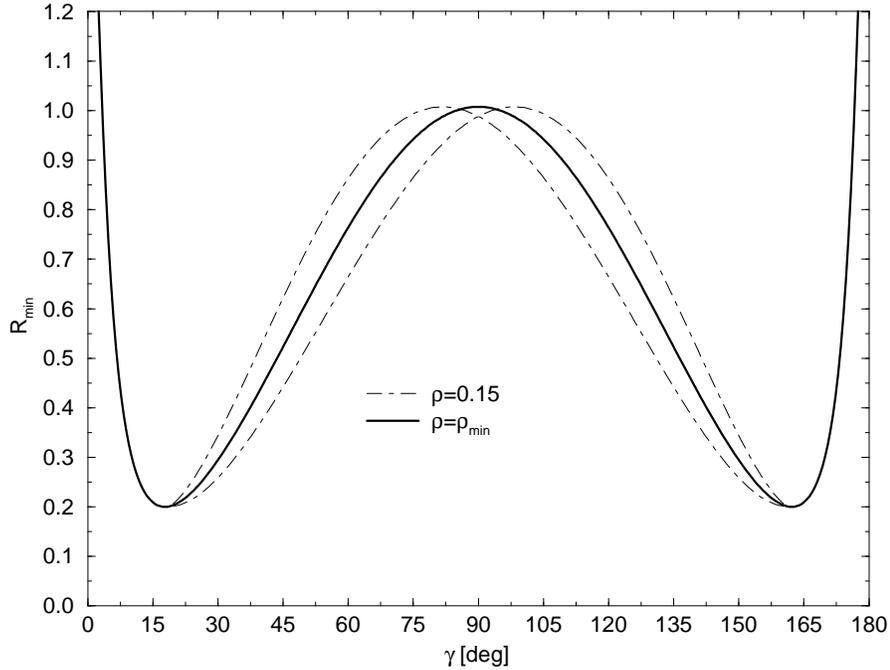}}}
\caption{The dependence of $R_{\rm min}$ on the CKM angle $\gamma$ 
for $|A_0|=0.2$ in the presence of rescattering effects, leading to
$|A_+|=0.1$ (electroweak penguins are neglected, i.e.\ 
$\epsilon=0$).}\label{fig:RminresAp}
\end{figure}

In Fig.~\ref{fig:RminresAp} we assume that the asymmetries $|A_+|=0.1$ 
and $|A_0|=0.2$ have been measured and show the dependence of $R_{\rm min}$
on $\gamma$ for $(\rho,\theta)=(\rho_{\rm min},\theta_0)$ and $\rho=0.15$.
If $\rho$ were known, we would have two solutions for $R_{\rm min}$. Assuming 
on the other hand that $|A_+|=0.1$ is due to $\rho\le0.15$, we get
an uncertainty of $\pm\,6^\circ$ for the bound on $\gamma$ corresponding to
$R=0.65$ and $|A_0|=0.2$. Furthermore, this figure illustrates nicely how 
values of $\gamma$ around $0^\circ$ and $180^\circ$ can be excluded 
through $|A_+|\not=0$, as we noted above. In the case of the dot-dashed 
lines, these values of $\gamma$ correspond to $\rho_{\rm min}>0.15$. 
Consequently, the allowed range
\begin{equation}
7^\circ\leq\gamma\leq53^\circ\quad\lor\quad127^\circ\leq\gamma\leq173^\circ,
\end{equation}
which would be implied by $R=0.65$ and $|A_0|=0.2$, is modified through 
rescattering effects with $\rho\le0.15$ -- leading to $|A_+|=0.1$ -- as 
follows:
\begin{equation}
20^\circ\leq\gamma\leq59^\circ\quad\lor\quad121^\circ\leq\gamma\leq160^\circ.
\end{equation}

\boldmath
\subsection{Including Rescattering Effects in the Bounds on $\gamma$ through
$B^\pm\to K^\pm K$}\label{FSI-incl}
\unboldmath
Concerning the bounds on $\gamma$ provided by (\ref{Rmin}), the rescattering
effects can be {\it included completely} by relating the decay 
$B^+\to\pi^+K^0$ to the mode $B^+\to K^+\overline{K^0}$ with the help of the 
$SU(3)$ flavour symmetry of strong interactions. Since these decays are 
actually related to each other by interchanging all $d$ and $s$ quarks, 
the so-called $U$ spin of the $SU(3)$ flavour symmetry suffices to this end. 

Using the unitarity of the CKM matrix and a notation similar to that in 
(\ref{Bpampl}), we get
\begin{equation}
A(B^+\to K^+\overline{K^0})=\lambda^3A\left[1-\left(
\frac{1-\lambda^2}{\lambda^2}\right)
\rho^{(d)}\,e^{i\theta_d}\,e^{i\gamma}\right]{\cal P}_{tc}^{(d)}\,,
\end{equation}
where
\begin{equation}\label{rhod-def}
\rho^{(d)}\,e^{i\theta_d}=\frac{\lambda^2R_b}{1-\lambda^2/2}
\left[1-
\left(\frac{{\cal P}_{uc}^{(d)}+{\cal A}^{(d)}}{{\cal P}_{tc}^{(d)}}
\right)\right]
\end{equation}
corresponds to (\ref{rho-def}). Consequently, direct CP violation in 
$B^+\to K^+\overline{K^0}$ is described in analogy to (\ref{Ap-def}) by
\begin{eqnarray}
\lefteqn{A_+^{(d)}\equiv\frac{\mbox{BR}(B^+\to K^+\overline{K^0})-
\mbox{BR}(B^-\to K^-K^0)}{\mbox{BR}(B^+\to K^+\overline{K^0})+
\mbox{BR}(B^-\to K^-K^0)}}\nonumber\\
&&~~~=\frac{2\,\lambda^2\,(1-\lambda^2)\,\rho^{(d)}
\sin\theta_d\,\sin\gamma}{\lambda^4-2\,\lambda^2\,(1-\lambda^2)\,\rho^{(d)}
\cos\theta_d\,\cos\gamma+(1-\lambda^2)^2\rho^{(d)\,2}}\,.\label{Apd-def}
\end{eqnarray}
As was pointed out in \cite{bfm}, another important quantity to deal with 
rescattering effects is the following ratio of combined branching ratios:
\begin{eqnarray}
\lefteqn{H\equiv R_{SU(3)}^{\,2}\left(\frac{1-\lambda^2}{\lambda^2}\right)
\frac{\mbox{BR}(B^\pm\to K^\pm K)}{\mbox{BR}(B^\pm\to\pi^\pm K)}}\nonumber\\
&&~=\frac{\lambda^4-2\,\lambda^2\,(1-\lambda^2)\,\rho^{(d)}\cos\theta_d\,
\cos\gamma+(1-\lambda^2)^2\rho^{(d)\,2}}{\lambda^4\left(1+
2\,\rho\,\cos\theta\,\cos\gamma+\rho^2\right)}\,,\label{H-def}
\end{eqnarray}
where BR$(B^\pm\to K^\pm K)$ is defined in analogy to (\ref{BR-char}),
tiny phase-space effects have been neglected (for a more detailed 
discussion, see \cite{fm2}), and 
\begin{equation}
R_{SU(3)}=\frac{M_B^2-M_\pi^2}{M_B^2-M_K^2}\,
\frac{F_{B\pi}(M_K^2;0^+)}{F_{BK}(M_K^2;0^+)}
\end{equation}
describes factorizable $SU(3)$ breaking. Here $F_{B\pi}(M_K^2;0^+)$ and 
$F_{BK}(M_K^2;0^+)$ are form factors parametrizing the hadronic quark-current
matrix elements 
$\langle\pi|(\bar b d)_{{\rm V-A}}|B\rangle$ and $\langle
K|(\bar b s)_{{\rm V-A}}|B\rangle$, respectively. Using, for example, the
model of Bauer, Stech and Wirbel \cite{BSW}, we have 
$R_{SU(3)}={\cal O}(0.7)$. At present, there is unfortunately no reliable 
approach available to deal with non-factorizable $SU(3)$ breaking. Since 
already the factorizable corrections are significant, we expect that 
non-factorizable $SU(3)$ breaking may also lead to sizeable effects. 

The three observables $H$, $A_+$ and $A_+^{(d)}$ depend on the four 
``unknowns'' $\rho$, $\theta$, $\rho^{(d)}$, $\theta_d$, and of course also
on the CKM angle $\gamma$. Using an additional $SU(3)$ input, either 
\begin{equation}\label{SU3-input}
\rho=\zeta_\rho\,\rho^{(d)}\,\,\quad\mbox{or}\quad\,\,\theta=\zeta_\theta\,
\theta_d\,,
\end{equation}
we are in a position to extract these quantities as functions of $\gamma$ 
from the measured values of $H$, $A_+$ and $A_+^{(d)}$, provided either 
$\zeta_\rho$ or $\zeta_\theta$, which parametrize $SU(3)$-breaking 
corrections, are known. As a first ``guess'', we may use $\zeta_\rho=1$ or 
$\zeta_\theta=1$. Keeping these $SU(3)$-breaking parameters explicitly in
our formulae, it is possible to study the sensitivity to deviations of 
$\zeta_{\rho,\theta}$ from 1, or to take into account $SU(3)$ breaking once 
we have a better understanding of this phenomenon. 

In order to include the rescattering effects in the bounds on $\gamma$ 
arising from (\ref{Rmin}), $\rho$ and $\theta$ determined this way are 
sufficient. The point is that we only have to know the dependence of 
$R_{\rm min}$ on $\gamma$ to constrain this CKM angle through the 
experimentally determined values of $R$ and $A_0$. The modification of 
this $\gamma$ dependence through rescattering effects can, however, be 
determined with the help of $\rho$ and $\theta$ obtained by using the 
approach discussed above. Consequently, the decays $B^+\to\pi^+K^0$ and 
$B^+\to K^+\overline{K^0}$ play a key role in taking into account final-state 
interactions in our bounds on $\gamma$. As we will see below, important 
by-products of this strategy are a range for $\rho$, and the exclusion of 
values of $\gamma$ within regions around $0^\circ$ and $180^\circ$. This 
approach works even in the case of ``trivial'' strong phases 
$\theta,\,\theta_d\in\{0^\circ,180^\circ\}$, where $B^+\to\pi^+K^0$ and 
$B^+\to K^+\overline{K^0}$ would exhibit no CP-violating effects.
 
It is interesting to note that a stronger $SU(3)$ input than 
(\ref{SU3-input}), assuming
\begin{equation}\label{SU3-input2}
\rho=\rho^{(d)}\,\,\quad\mbox{and}\quad\,\,\theta=\theta_d\,,
\end{equation}
yields a nice relation between $A_+$, $A_+^{(d)}$ and the combined 
$B^\pm\to\pi^\pm K$ and $B^\pm\to K^\pm K$ branching ratios:
\begin{equation}\label{ApApdrel}
\frac{A_+}{A_+^{(d)}}=-\,R_{SU(3)}^{\,2}\,\frac{\mbox{BR}(B^\pm\to K^\pm 
K)}{\mbox{BR}(B^\pm\to\pi^\pm K)}=-\left(\frac{\lambda^2}{1-\lambda^2}
\right)H\,,
\end{equation}
which has already been pointed out in \cite{bfm}. This expression implies
opposite signs for $A_+$ and $A_+^{(d)}$ and, moreover, allows a 
determination of $H$ and of the $SU(3)$-breaking parameter $R_{SU(3)}$. 
A future experiment finding that $A_+$ and $A_+^{(d)}$ have equal signs 
would mean either that $\sin\theta$ and $\sin\theta_d$ have opposite signs, 
or contributions from ``new physics''. 

The present upper limit from the CLEO collaboration \cite{cleo} on the 
combined $B^\pm\to K^\pm K$ branching ratio is given by 
BR$(B^\pm\to K^\pm K)<2.1\times10^{-5}$. Let us have a closer look at the 
impact of rescattering effects on $B^\pm\to K^\pm K$. To this 
end we assume $R_{SU(3)}=0.7$, the $SU(3)$ relations given in 
(\ref{SU3-input2}), $\gamma=50^\circ$, $\theta=25^\circ$, and 
BR$(B^\pm\to\pi^\pm K)=2.3\times10^{-5}$, which is the central 
value of present CLEO data. In order to discuss the
case of tiny rescattering effects, we use $\rho=0.02$, yielding
BR$(B^\pm\to K^\pm K)=1.6\times10^{-6}$, \mbox{$A_+^{(d)}=+\,37\%$} and
$A_+=-\,1.3\%$. These values are in accordance with the results 
obtained by performing model calculations at the perturbative quark level 
\cite{pert-pens}. In this case, the CLEO bound on BR$(B^\pm\to K^\pm K)$ 
would be one order of magnitude above the estimated branching
ratio. However, in contrast to $B^\pm\to\pi^\pm K$, where only the CP asymmetry
may be enhanced sizeably through final-state interactions related to
(\ref{u-res}) and the branching ratio remains essentially unchanged, the 
decay rate for $B^\pm\to K^\pm K$ may be affected dramatically by such
rescattering processes. To illustrate this remarkable feature, we use 
$\rho=0.15$, yielding BR$(B^\pm\to K^\pm K)=1.2\times10^{-5}$, 
$A_+^{(d)}=+\,30\%$ and $A_+=-\,8.1\%$. While the rescattering effects 
lead to some reduction of $A_+^{(d)}$, in this example, they enhance the 
branching ratio for $B^\pm\to K^\pm K$ by almost a factor 10, and could 
thereby make an experimental study of this mode feasible. These 
considerations demonstrate that BR$(B^\pm\to K^\pm K)$ may actually be much 
closer to the present CLEO bound than expected from simple quark-level 
estimates, if rescattering effects are in fact large. Consequently, 
$B^\pm\to K^\pm K$ may be a very promising mode not just to constrain, 
but to {\it control} rescattering effects in the bounds on the CKM angle 
$\gamma$ arising from $(\ref{Rmin})$.

\begin{figure}
\centerline{
\rotate[r]{
\epsfxsize=9.2truecm
\epsffile{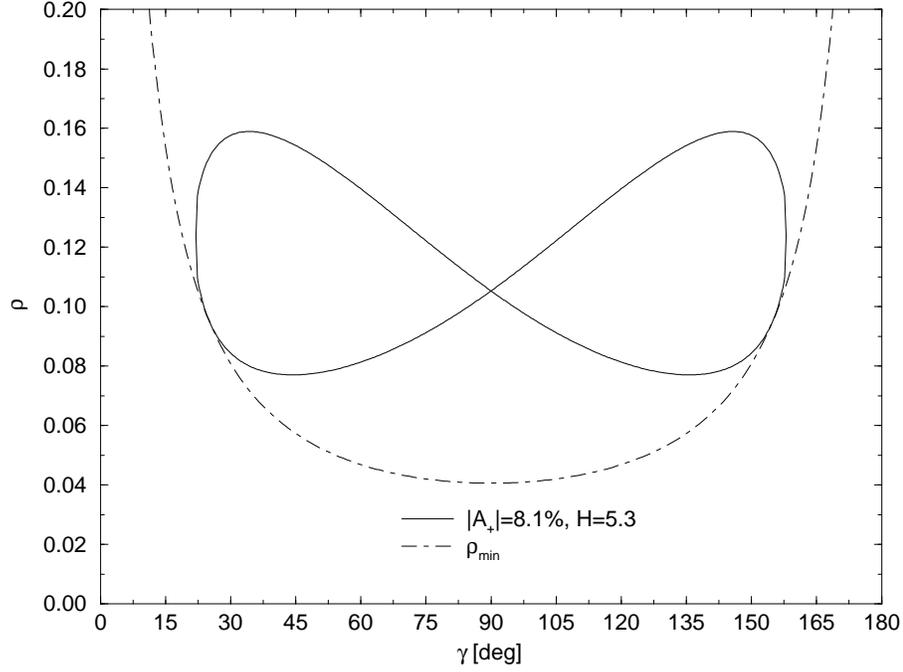}}}
\caption{The dependence of $\rho$ on $\gamma$ obtained by relating 
$B^\pm\to\pi^\pm K$ and $B^\pm\to K^\pm K$ through the $SU(3)$ flavour
symmetry for the example discussed in the text.}\label{fig:Bkkrho}
\end{figure}

\begin{figure}
\centerline{
\rotate[r]{
\epsfxsize=9.2truecm
\epsffile{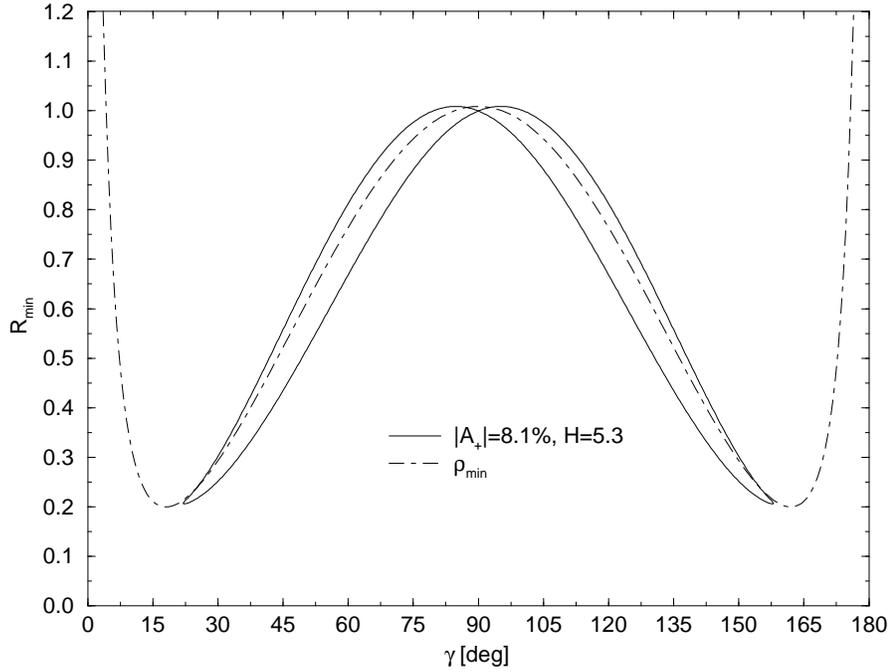}}}
\caption{The dependence of $R_{\rm min}$ on $\gamma$ obtained by relating 
$B^\pm\to\pi^\pm K$ and $B^\pm\to K^\pm K$ through the $SU(3)$ flavour 
symmetry for the example discussed in the text 
($|A_0|=0.2$).}\label{fig:BkkRmin}
\end{figure}

In order to put this statement on a more quantitative ground, let us assume 
that future experiments find BR$(B^\pm\to\pi^\pm K)=2.3\times10^{-5}$, 
$A_+=-\,8.1\%$ and BR$(B^\pm\to K^\pm K)=1.2\times10^{-5}$, 
$A_+^{(d)}=+\,30\%$, just as in the example considered above. To simplify our 
discussion, let us again use the $SU(3)$ relations listed in
(\ref{SU3-input2}). With the help of (\ref{ApApdrel}) we would then obtain
$R_{SU(3)}=0.7$ and $H=5.3$, and the quantitiy $w$ containing essentially all 
the information needed to take into account the rescattering effects in 
(\ref{Rmin}) is given by
\begin{equation}\label{w-deter}
w=\frac{1}{\lambda}\,\sqrt{\frac{\rho^2+\lambda^2\,(1-\rho^2)}{1+
\lambda^2\,(H-1)}}\,.
\end{equation}
Moreover we obtain a simple expression for $\cos\theta$. Combining it with 
(\ref{Theta-deter}) yields a quadratic equation for $\rho^2$, i.e.\ we
can fix this parameter up to a two-fold ambiguity. Using (\ref{w-deter}),
$w$ can hence be determined up to a two-fold ambiguity as well. In 
Fig.~\ref{fig:Bkkrho} we show the dependence of $\rho$ extracted this way
on the CKM angle $\gamma$ in the case of our example. We observe two 
interesting features. First, the range for $\rho$ is quite small in this 
case: $0.08\le\rho\le0.16$. Secondly, values of $\gamma$ within 
$0^\circ\le\gamma\le22^\circ$ $\lor$ $158^\circ\le\gamma\le 180^\circ$ 
are excluded. Assuming that $A_0=\pm\,0.2$ has been measured and using 
(\ref{Rmin-simplFSI}) and (\ref{w-deter}), we obtain the solid lines shown 
in Fig.~\ref{fig:BkkRmin}, where we have also included the curve 
corresponding to $(\rho,\theta)=(\rho_{\rm min},\theta_0)$. For a given
value of $R$, we can read off easily the allowed range for $\gamma$
taking into account the rescattering effects. In the case of $R=0.65$,
we would obtain $22^\circ\leq\gamma\leq59^\circ$ $\lor$ 
$121^\circ\leq\gamma\leq158^\circ$.

\boldmath
\subsection{Rescattering Effects in Strategies to Determine 
$\gamma$}\label{FSI-r}
\unboldmath
In order not just to constrain, but to extract the CKM angle $\gamma$ from 
the $B\to\pi K$ decays considered in this paper, information on $r$ 
is essential, as we have seen in Section~\ref{gamma-det}. Before we focus
on this quantity, let us first have a closer look at the modification of 
the contours in the $\gamma$--$r$ plane (see Fig.~\ref{fig:r-det065})
through rescattering effects. 

\begin{figure}
\centerline{
\rotate[r]{
\epsfxsize=9.2truecm
\epsffile{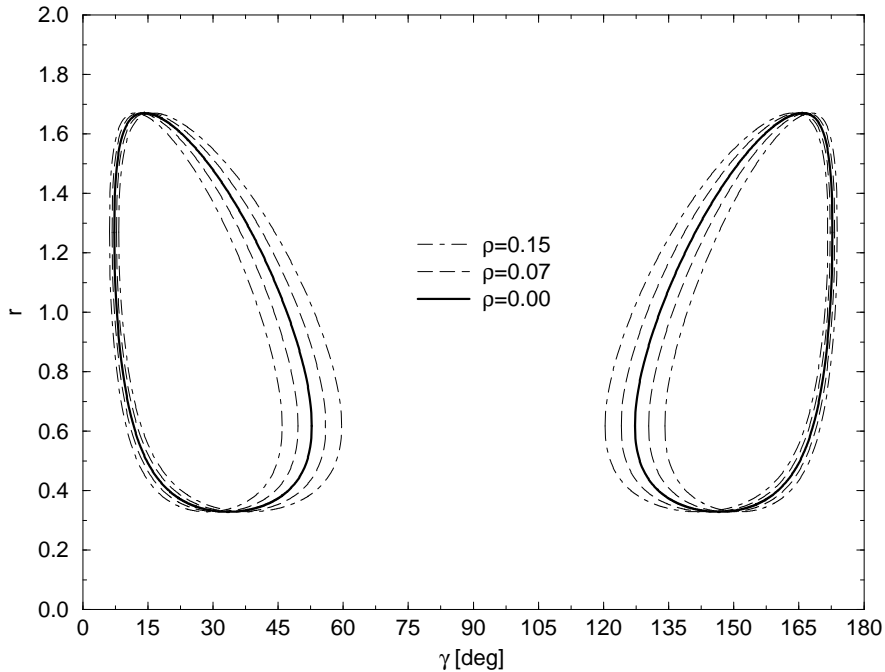}}}
\caption{The shift of the contours in the $\gamma$--$r$ plane corresponding
to $R=0.65$ and $|A_0|=0.2$ through rescattering effects ($\theta
\in\{0^\circ,180^\circ\}$, $\epsilon=0$).}\label{fig:r-det065res}
\end{figure}

\begin{figure}
\centerline{
\rotate[r]{
\epsfxsize=9.2truecm
\epsffile{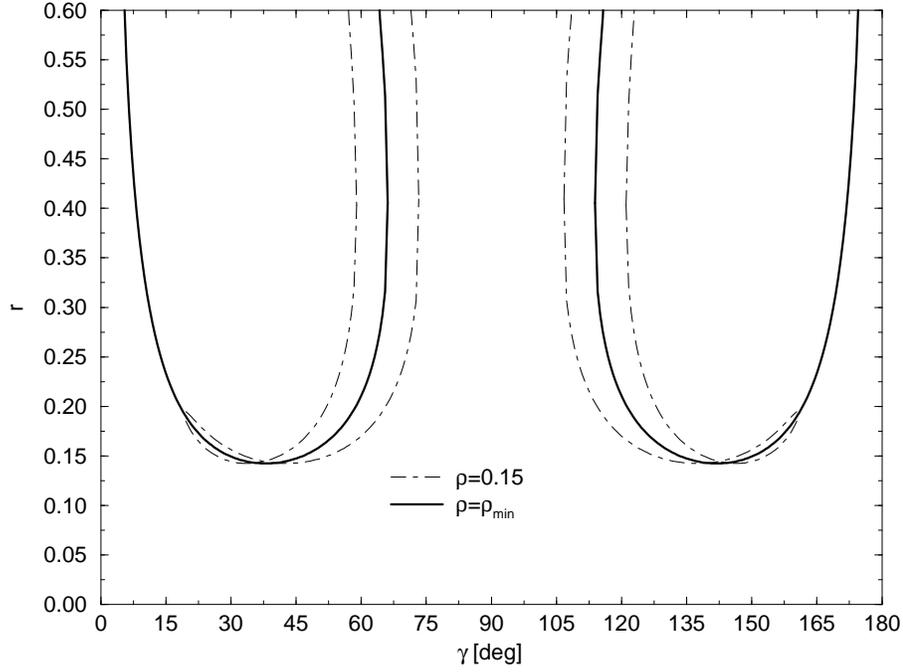}}}
\caption{Contours in the $\gamma$--$r$ plane corresponding to $R=0.85$, 
$A_0=\pm\,0.2$ and $A_+=\pm\,0.1$ for neglected electroweak penguin
contributions, i.e.\ $\epsilon=0$.}\label{fig:r-det065res01p}
\end{figure}

\begin{figure}
\centerline{
\rotate[r]{
\epsfxsize=9.2truecm
\epsffile{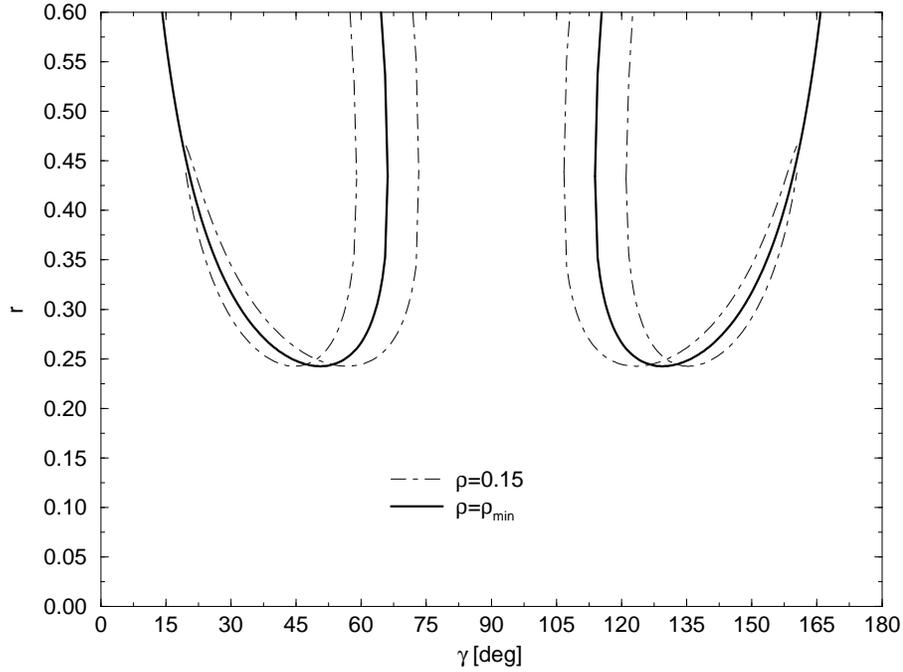}}}
\caption{Contours in the $\gamma$--$r$ plane corresponding
to $R=0.85$, $A_0=\pm\,0.2$ and $A_+=\mp\,0.1$ for neglected electroweak 
penguin contributions, i.e.\ $\epsilon=0$.}\label{fig:r-det065res01m}
\end{figure}

In Fig.~\ref{fig:r-det065res} we show the shift of the contours corresponding
to $R=0.65$ and $|A_0|=0.2$ for $\theta\in\{0^\circ,180^\circ\}$. In this 
case, $B^+\to\pi^+K^0$ would exhibit no CP violation. On the other hand,
the situation arising if a non-vanishing value of $A_+$ has been measured is
illustrated in Figs.~\ref{fig:r-det065res01p} and \ref{fig:r-det065res01m} 
for $A_0=\pm\,0.2$, $A_+=\pm\,0.1$ and $A_0=\pm\,0.2$, $A_+=\mp\,0.1$, 
respectively. There we have chosen $R=0.85$, favouring smaller values of $r$
than $R=0.65$, i.e.\ values that are closer to the ``factorized'' result 
(\ref{r-fact}). We observe that there is an interesting difference between 
the cases where the CP asymmetries $A_0$ and $A_+$ have equal or opposite
signs. In the former case even lower values of $r$ are favoured. If we could 
fix $r$, we would be in a position to extract the CKM angle $\gamma$ up to 
discrete ambiguities with the help of these figures. In this example, the 
uncertainty of $\gamma$ would be at most $\pm\,8^\circ$, if we assume that 
the CP asymmetries $A_+$ are due to $\rho\le0.15$. Such an assumption can be 
avoided, if we apply the $B^\pm\to K^\pm K$ approach outlined in 
Subsection~\ref{FSI-incl}, which allows us to take into account the 
rescattering effects also in the contours in the $\gamma$--$r$ plane.

Unfortunately, the value of $r$ is also affected by rescattering processes
and it is not possible to include them in a similarly ``easy'' way as in 
the case of these contours. In order to discuss this subtle point,
let us have a closer look at the amplitude $T$ defined by (\ref{T-def}).
Using the low-energy effective Hamiltonian describing $B^+\to\pi^+K^0$,
$B^0_d\to\pi^-K^+$ decays, which takes the form \cite{bbl}
\begin{equation}\label{heff}
{\cal H}_{{\rm eff}} = \frac{G_{\rm F}}{\sqrt{2}}\left[ 
\lambda_u^{(s)}\sum_{k=1}^2C_k(\mu) Q_k^u +\lambda_c^{(s)}
\sum_{k=1}^2C_k(\mu) Q_k^c -\lambda_t^{(s)} \sum^{10}_{k=3} 
C_k(\mu) Q_k \right],
\end{equation}
where $Q_3$,\ldots,$Q_6$ and $Q_7$,\ldots,$Q_{10}$ are QCD and electroweak 
penguin operators, respectively, and where $\mu={\cal O}(m_b)$ is a 
renormalization scale, as well as the isospin symmetry of strong interactions, 
$T$ can be expressed in terms of hadronic matrix elements of the 
current--current operators (\ref{CC-ops}) as follows \cite{bfm}:
\begin{eqnarray}
\lefteqn{T\equiv-\,\frac{G_{\rm F}}{\sqrt{2}}\,\lambda^4\,A\,R_b\biggl[
C_1(\mu) \langle K^+\pi^-|Q_1^u(\mu)|B^0_d\rangle_{\rm T}+
C_2(\mu) \langle K^+\pi^-|Q_2^u(\mu)|B^0_d\rangle_{\rm T}~~}\nonumber\\
&&~~~~~~~~+\Bigl\{C_1(\mu) \langle K^+\pi^-|Q_1^u(\mu)|B^0_d\rangle_{\rm P} 
+C_2(\mu) \langle K^+\pi^-|Q_2^u(\mu)|B^0_d\rangle_{\rm P}~~\nonumber\\
&&~~~~~~~~-\,C_1(\mu) \langle K^+\pi^-|Q_1^d(\mu)|B^0_d\rangle
-\,C_2(\mu) \langle K^+\pi^-|Q_2^d(\mu)|B^0_d\rangle\Bigr\}
\biggr]\,e^{i\gamma},\label{T-opex}
\end{eqnarray}
where we have to perform the replacement $u\to d$ in (\ref{CC-ops}) in order
to get the expressions for $Q^d_{1,2}$. The labels ``T'' and ``P'' denote
insertions of the current--current operators $Q^u_{1,2}$ into tree-diagram-like
and penguin-like topologies. While the terms in (\ref{T-opex}) with 
label ``T'' correspond to the $\tilde{\cal T}$ amplitude in (\ref{T-def}), 
the $Q_{1,2}^d$ contributions in the term in curly brackets are required in 
order to apply the $SU(2)$ isospin symmetry correctly to relate the decays 
$B^+\to\pi^+K^0$ and $B^0_d\to\pi^-K^+$ \cite{bfm}. The insertions of 
$Q^u_{1,2}$ into penguin-like topologies in this term correspond to the 
$\tilde P_u$ amplitude in (\ref{T-def}), while the $Q_{1,2}^d$ operators 
contribute both through insertions into penguin topologies and through 
annihilation processes and describe the combination ${\cal A}+P_u$ in 
(\ref{T-def}). The electroweak penguin amplitudes appearing in that 
expression are neglected in (\ref{T-opex}). They play a minor role for 
$T$, as we will see in the next section, where the operator expression 
we shall give for $T$ will take into account also electroweak penguins.

\begin{figure}
\begin{center}
\leavevmode
\vspace*{1truecm} 
\rotate[r]{
\epsfysize=11.8truecm 
\epsffile{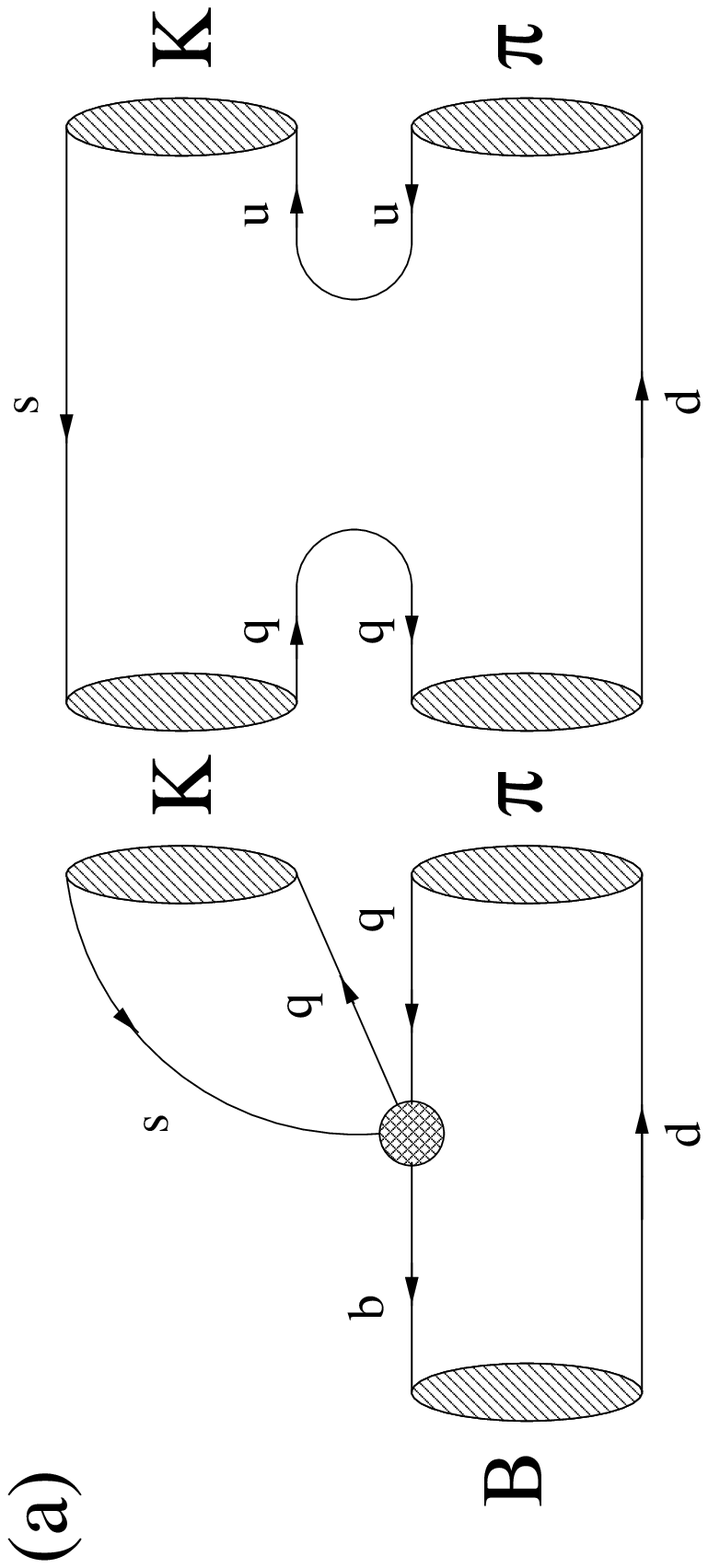}}
\vspace*{1truecm} 
\rotate[r]{
\epsfysize=11.5truecm 
\epsffile{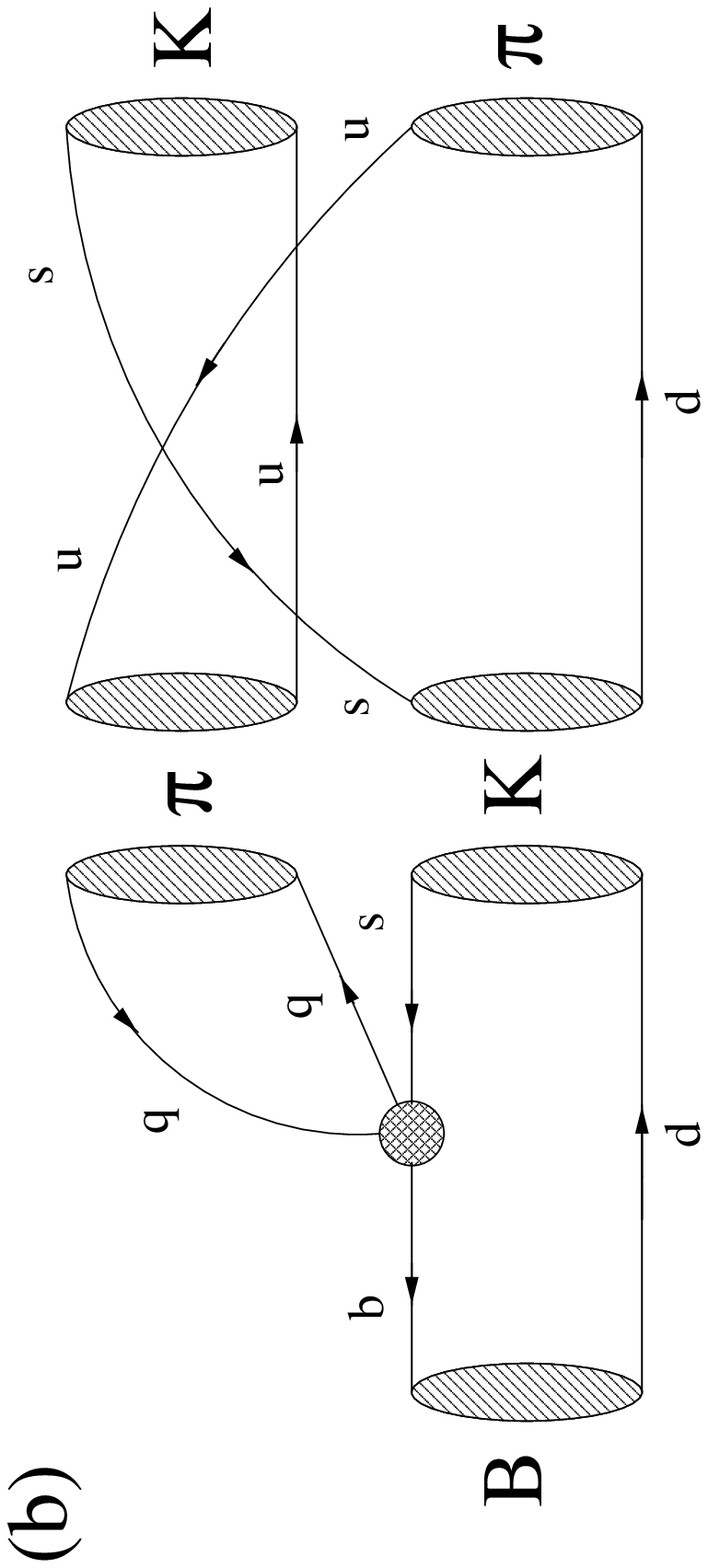}}
\rotate[r]{
\epsfysize=11.3truecm 
\epsffile{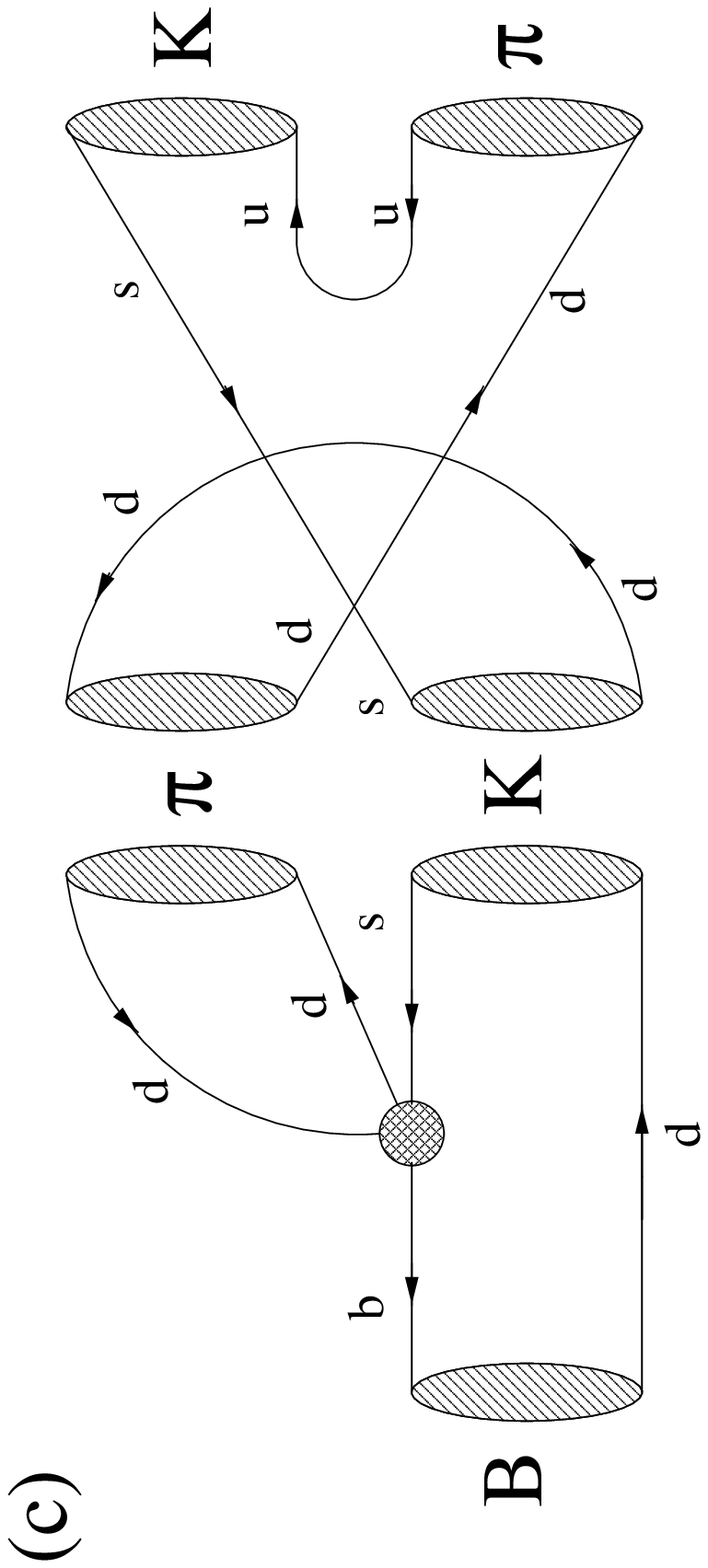}}
\end{center}
\caption{Contributions to the amplitude $T$ through rescattering processes 
of the kind $B^0_d\to\{\pi^-K^+\}\to\pi^-K^+$ and 
$B^0_d\to\{\pi^0K^0\}\to\pi^-K^+$. The shaded circles represent insertions
of the current--current operators $Q_{1,2}^q$, where $q\in\{u,d\}$. In the 
case of the annihilation topology (c), only the $Q_{1,2}^d$ operators 
contribute.}\label{fig:rescatter-topol}
\end{figure}
 
While the short-distance contributions to the term in curly brackets in
(\ref{T-opex}) cancel, this is not the case for the long-distance 
contributions associated with final-state interactions. In 
Fig.~\ref{fig:rescatter-topol} we show some of the corresponding Feynman 
diagrams, where (a) and (b) represent insertions of the current--current 
operators $Q_{1,2}^q$ ($q\in\{u,d\}$) into penguin-like topologies, whereas 
(c) is an annihilation topology involving only $Q_{1,2}^d$. Having a look 
at these diagrams, it is obvious that their contributions do not in general 
cancel in (\ref{T-opex}). Concerning the topologies (a), the $Q_{1,2}^u$
operators contribute through rescattering processes of the type 
$B^0_d\to\{\pi^-K^+\}\to\pi^-K^+$, while the $Q_{1,2}^d$ operators contribute
through $B^0_d\to\{\pi^0K^0\}\to\pi^-K^+$ and involve the $\bar d\,d/
\sqrt{2}$ piece of a neutral pion. In the latter case, also 
$B^0_d\to\{\eta\,K^0,\ldots\}\to\pi^-K^+$ processes are expected to play an 
important role. In the case of the topologies (b), only rescattering 
processes of the kind $B^0_d\to\{\pi^0K^0\}\to\pi^-K^+$ contribute. Since 
$\bar u\,u$ and $\bar d\,d$ enter in the $\pi^0$ wave function with opposite 
signs, the topologies (b) with $Q_{1,2}^u$ and $Q_{1,2}^d$ insertions have 
opposite signs as well and do not cancel in the difference in (\ref{T-opex}). 
The annihilation topology (c) has anyway no counterpart with $Q_{1,2}^u$ 
insertions. Consequently, the rescattering contributions to the term in curly 
brackets in (\ref{T-opex}) do not cancel. If they are of the same order of 
magnitude as $\tilde{\cal T}$ describing the strength of the ``first step'' 
in the rescattering processes of the type (a) -- such a scenario is found in 
the model calculation performed in \cite{fknp} -- the value of $r$ could in 
principle be shifted significantly from its ``factorized'' value 
(\ref{r-fact}). 

The feature described in the previous paragraph provides an interesting 
mechanism to generate values of $r$ larger than those obtained within 
the framework of ``factorization'', and may be the reason for the fact that 
the values of $r$ preferred by present CLEO data are at the edge of 
compatibility with (\ref{r-fact}), as we noted in 
Subsection~\ref{r-bounds}. In particular, the small central value of 
$R=0.65$ may indicate already that $r$ is enhanced considerably by 
final-state interactions, and it may well be possible that future 
measurements will stabilize around this na\"\i vely small value. 

Although this is good news for the bounds on $\gamma$ from $B\to\pi K$
decays, it is bad news for the corresponding extractions of this CKM angle, 
which require the knowledge of $r$. Unfortunately, in the case of this 
quantity, final-state interactions cannot be taken into account in a  
simple way. Consequently, expectations based on factorization that a future 
theoretical uncertainty of $r$ as small as ${\cal O}(10\%)$ may be achievable 
\cite{groro,wuegai} appear too optimistic. If we look, however, at 
Figs.~\ref{fig:r-det065res01p} and \ref{fig:r-det065res01m}, we observe
that the dependence of $\gamma$ on $r$ is very weak for 
$r\mathrel{\hbox{\rlap{\hbox{\lower4pt\hbox{$\sim$}}}\hbox{$>$}}}0.3$
in this example; even for values of $R$ as large as 0.85, a significant
region around $\gamma=90^\circ$ could be excluded. The power of a future
accurate measurement of the decays $B^\pm\to\pi^\pm K$ and $B_d\to\pi^\mp
K^\pm$ is therefore probably not a ``precision'' measurement of $\gamma$, 
but phenomenologically interesting constraints on this CKM angle. 

\boldmath
\section{The Role of Electroweak Penguins}\label{EWP-effects}
\begin{figure}
\centerline{
\rotate[r]{
\epsfxsize=9.2truecm
\epsffile{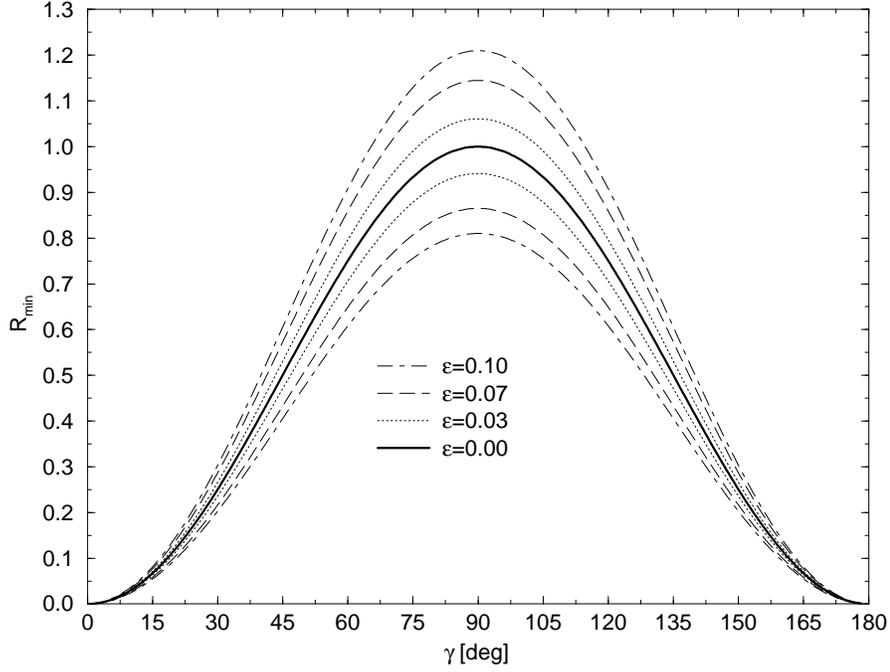}}}
\caption{The effect of electroweak penguins on $R_{\rm min}$ for $A_0=0$.
The curves for a given value of $\epsilon$ correspond to 
$\Delta\in\{0^\circ,180^\circ\}$ and represent the maximal shift from 
$\epsilon=0$.}\label{fig:Rminewp0}
\end{figure}

\begin{figure}
\centerline{
\rotate[r]{
\epsfxsize=9.2truecm
\epsffile{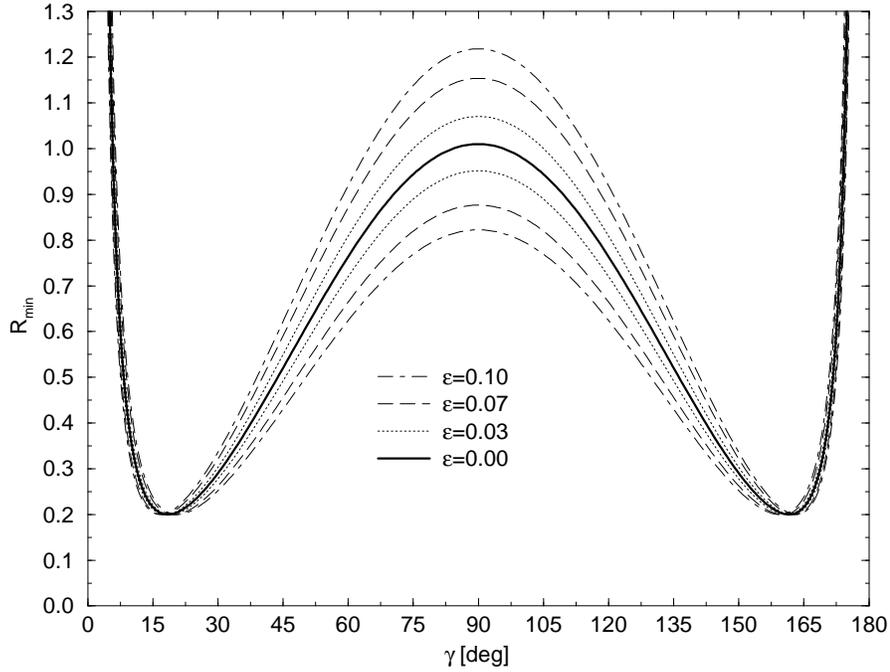}}}
\caption{The effect of electroweak penguins on $R_{\rm min}$ for $|A_0|=0.2$.
The curves for a given value of $\epsilon$ correspond to 
$\Delta\in\{0^\circ,180^\circ\}$ and represent the maximal shift from 
$\epsilon=0$.}\label{fig:RminewpA0}
\end{figure}

\unboldmath
Let us begin our discussion of electroweak penguin effects by first having 
a look at (\ref{Rmin}). For $\rho=0$, i.e.\ neglected rescattering effects,
the modification of $R_{\rm min}$ through electroweak penguins is described
by $\kappa=1+2\,\epsilon\,\cos\Delta+\epsilon^2$. These effects are minimal
and only of second order in $\epsilon$ for $\Delta\in\{90^\circ,270^\circ\}$, 
and maximal for $\Delta\in\{0^\circ,180^\circ\}$. In the case of 
$\Delta=0^\circ$, the bounds on $\gamma$ get stronger, excluding a larger
region around $\gamma=90^\circ$, while they are weakened for 
$\Delta=180^\circ$. In Figs.~\ref{fig:Rminewp0} and \ref{fig:RminewpA0}
we show the maximal electroweak penguin effects for various values of 
$\epsilon$. The electroweak penguins are ``colour-suppressed'' in the case 
of $B^+\to\pi^+K^0$ and $B^0_d\to\pi^-K^+$, and estimates based on simple 
calculations performed at the perturbative quark level, where
the relevant hadronic matrix elements are treated within the ``factorization''
approach, typically give $\epsilon={\cal O}(1\%)$ \cite{fm2}. Even for such
small values of $\epsilon$, there is a sizeable shift of $R_{\rm min}$, as
can be seen in Figs.~\ref{fig:Rminewp0} and \ref{fig:RminewpA0}. Comparing 
these figures with Fig.~\ref{fig:Rminres0}, we observe that $R_{\rm min}$
is more sensitive to electroweak penguin than to rescattering effects. 
Since the crude estimates yielding $\epsilon={\cal O}(1\%)$ may well 
underestimate the role of electroweak penguins \cite{groro,neubert,fm3}, an
improved theoretical description of these topologies is highly desirable.

\boldmath
\subsection{An Improved Theoretical Description}\label{EWPs-impr}
\unboldmath

The relevant electroweak penguin amplitude affecting bounds and extractions
of the CKM angle $\gamma$ from $B^\pm\to\pi^\pm K$, $B_d\to\pi^\mp K^\pm$
decays has been given in (\ref{Pew-def}), where $P_{\rm ew}^t$ and
$\tilde P_{\rm ew}^t$ correspond to electroweak penguins with internal
top quarks and can be expressed in terms of hadronic matrix elements of 
four-quark operators as follows (see (\ref{heff})):
\begin{eqnarray}
P_{\rm ew}^t&=&-\,\frac{G_{\rm F}}{\sqrt{2}}\sum_{k=7}^{10}
C_k(\mu)\langle K^0\pi^+|Q_k(\mu)|B^+\rangle\\
\tilde P_{\rm ew}^t&=&\frac{G_{\rm F}}{\sqrt{2}}\sum_{k=7}^{10}
C_k(\mu)\langle K^+\pi^-|Q_k(\mu)|B^0_d\rangle.
\end{eqnarray}
The electroweak penguin operators $Q_7$,\ldots,$Q_{10}$ have the
following structure:
\begin{equation}
Q_k=\frac{3}{2}\sum_{q=u,d,c,s,b}c_q\, Q_k^q\,,
\end{equation}
where $c_q$ denotes the electrical quark charges and the four-quark operators
are given by 
\begin{equation}
Q_7^q = (\bar b_{\alpha} s_{\alpha})_{{\rm V-A}}(\bar 
q_{\beta}q_{\beta})_{{\rm V+A}}\,,\quad
Q_8^q = (\bar b_{\alpha} s_{\beta})_{{\rm V-A}}(\bar 
q_{\beta} q_{\alpha})_{{\rm V+A}} 
\end{equation}
\begin{equation}
~~Q_9^q = (\bar b_{\alpha} s_{\alpha})_{{\rm V-A}}(\bar 
q_{\beta}q_{\beta})_{{\rm V-A}}\,,\quad
Q_{10}^q = (\bar b_{\alpha} s_{\beta})_{{\rm V-A}}(\bar 
q_{\beta} q_{\alpha})_{{\rm V-A}}\,. 
\end{equation}
The amplitude $P_{\rm ew}^t$ can therefore be written as
\begin{eqnarray}
\lefteqn{P_{\rm ew}^t=-\,\frac{G_{\rm F}}{\sqrt{2}}\,\frac{3}{2}
\left[c_d\sum_{k=7}^{10}
C_k(\mu)\langle K^0\pi^+|Q_k^d(\mu)|B^+\rangle_{\rm T}+c_d\sum_{k=7}^{10}
C_k(\mu)\langle K^0\pi^+|Q_k^d(\mu)|B^+\rangle_{\rm P}~~\right.}\nonumber\\
&&\left.+\,c_u\sum_{k=7}^{10}C_k(\mu)\langle K^0\pi^+|Q_k^u(\mu)|B^+\rangle\,+
\sum_{q=c,s,b}c_q\left\{\sum_{k=7}^{10}C_k(\mu)
\langle K^0\pi^+|Q_k^q(\mu)|B^+\rangle\right\}\right].
\end{eqnarray}
The notation in this expression is as in (\ref{T-opex}). While the first
term describes the contributions of $Q_k^d$ arising from insertions into
tree-diagram-like topologies, the other terms contain in particular also
rescattering effects, which may play an important role. Applying the
$SU(2)$ isospin symmetry to the hadronic matrix elements of the $Q_k^q$
operators, we obtain
\begin{eqnarray}
\lefteqn{P_{\rm ew}^t=\frac{G_{\rm F}}{\sqrt{2}}\,\frac{3}{2}
\left[c_d\sum_{k=7}^{10}
C_k(\mu)\langle K^+\pi^-|Q_k^u(\mu)|B^0_d\rangle_{\rm T}+c_d\sum_{k=7}^{10}
C_k(\mu)\langle K^+\pi^-|Q_k^u(\mu)|B^0_d\rangle_{\rm P}~~\right.}\nonumber\\
&&\left.+\,c_u\sum_{k=7}^{10}C_k(\mu)\langle K^+\pi^-|Q_k^d(\mu)|
B^0_d\rangle\,+\sum_{q=c,s,b}c_q\left\{\sum_{k=7}^{10}C_k(\mu)
\langle K^+\pi^-|Q_k^q(\mu)|B^0_d\rangle\right\}\right]\label{EWP1},
\end{eqnarray}
while we have on the other hand
\begin{eqnarray}
\lefteqn{\tilde P_{\rm ew}^t=\frac{G_{\rm F}}{\sqrt{2}}\,\frac{3}{2}
\left[c_u\sum_{k=7}^{10}
C_k(\mu)\langle K^+\pi^-|Q_k^u(\mu)|B^0_d\rangle_{\rm T}+c_u\sum_{k=7}^{10}
C_k(\mu)\langle K^+\pi^-|Q_k^u(\mu)|B^0_d\rangle_{\rm P}~~\right.}\nonumber\\
&&\left.+\,c_d\sum_{k=7}^{10}C_k(\mu)\langle K^+\pi^-|Q_k^d(\mu)|
B^0_d\rangle\,+\sum_{q=c,s,b}c_q\left\{\sum_{k=7}^{10}C_k(\mu)
\langle K^+\pi^-|Q_k^q(\mu)|B^0_d\rangle\right\}\right].\label{EWP2}
\end{eqnarray}
Concerning the quantity $P_{\rm ew}$, only the difference of these 
amplitudes is relevant, which is given by
\begin{eqnarray}
\lefteqn{\tilde P_{\rm ew}^t-P_{\rm ew}^t=\left(c_u-c_d\right)
\frac{G_{\rm F}}{\sqrt{2}}\,\frac{3}{2}\left[\sum_{k=7}^{10}C_k(\mu)
\langle K^+\pi^-|Q_k^u(\mu)|B^0_d\rangle_{\rm T}\right.}\nonumber\\
&&\left.+\,\sum_{k=7}^{10}C_k(\mu)\left\{\langle K^+\pi^-|Q_k^u(\mu)|B^0_d
\rangle_{\rm P}-\langle K^+\pi^-|Q_k^d(\mu)|B^0_d\rangle\right\}
\right].\label{ampl-diff}
\end{eqnarray}
This expression is much simpler than (\ref{EWP1}) and (\ref{EWP2}), since
the contributions of $Q_k^q$ with $q=c,s,b$ -- including also rescattering 
contributions that are very hard to estimate -- cancel fortunately because of 
the isospin symmetry. Since only the Wilson coefficients $C_9$ and $C_{10}$ 
are sizeable \cite{bbl}, where $C_9$ plays the most important role and is 
about three times larger than $C_{10}$, the amplitude difference 
(\ref{ampl-diff}) simplifies further and we have only to care about
$Q_{9,10}^u$ and $Q_{9,10}^d$. If we compare these operators with the
current--current operators given in (\ref{CC-ops}), we observe that
they are related to each other through a simple Fierz transformation:
\begin{equation}
\left.Q_9^{u,d}\right|_{\rm Fierz}=\,Q_1^{u,d},\quad
\left.Q_{10}^{u,d}\right|_{\rm Fierz}=\,Q_2^{u,d}.
\end{equation}
Beyond the leading order, one has to be careful, when performing such Fierz 
transformations, since the next-to-leading order Wilson coefficients depend 
also on the chosen operator basis \cite{bbl}. Here we will only use
leading-order Wilson coefficients for simplicity, and will comment on
the influence of next-to-leading order corrections on our ``final'' result
below. 

The amplitudes $P_{\rm ew}^c$ and $\tilde P_{\rm ew}^c$ in (\ref{Pew-def}) 
corresponding to electroweak penguins with internal charm quarks, which 
would be needed for a consistent use of next-to-leading order Wilson 
coefficients \cite{bbl}, play a negligible role. The point is that 
electroweak penguins only become important because of the large top-quark 
mass, which compensates their suppression relative to the QCD penguins due 
to the small ratio $\alpha/\alpha_s={\cal O}(10^{-2})$ of the QED and QCD 
couplings. This feature is reflected by the sizeable value of $C_9$. 

Consequently we arrive at the following expression for $P_{\rm ew}$:
\begin{eqnarray}
P_{\rm ew}&=&-\lambda^2A\,\frac{G_{\rm F}}{\sqrt{2}}\,
\frac{3}{2}\,\biggl[C_9(\mu)\langle K^+\pi^-|
Q_1^u(\mu)|B^0_d\rangle_{\rm T}+C_{10}(\mu)\langle K^+\pi^-|Q_2^u(\mu)|
B^0_d\rangle_{\rm T}\nonumber\\
&&~~+\left\{C_9(\mu)\langle K^+\pi^-|Q_1^u(\mu)|B^0_d\rangle_{\rm P}
+C_{10}(\mu)\langle K^+\pi^-|Q_2^u(\mu)|B^0_d\rangle_{\rm P}\right.
\nonumber\\
&&\left.~~-C_9(\mu)\langle K^+\pi^-|Q_1^d(\mu)|B^0_d\rangle-C_{10}(\mu)
\langle K^+\pi^-|Q_2^d(\mu)|B^0_d\rangle\right\}\biggr],\label{Pew-opex}
\end{eqnarray}
where we have neglected the ${\cal O}(\lambda^4)$ term in (\ref{Pew-def})
and have used in addition $c_u-c_d=1$. Remarkably, (\ref{Pew-opex})
is completely analogous to the operator expression for $T$ given in
(\ref{T-opex}). Introducing the following non-perturbative ``bag'' parameters:
\begin{eqnarray}
\lefteqn{\frac{1}{3}B_1(\mu)
\langle K^+\pi^-|Q_2^u|B^0_d\rangle_{\rm T}^{\rm fact}}\nonumber\\
&&~~~~\equiv\langle K^+\pi^-|Q_1^u(\mu)|B^0_d\rangle_{\rm T}+\left\{
\langle K^+\pi^-|Q_1^u(\mu)|B^0_d\rangle_{\rm P}-
\langle K^+\pi^-|Q_1^d(\mu)|B^0_d\rangle\right\}\label{B1-def}
\end{eqnarray}
\begin{eqnarray}
\lefteqn{B_2(\mu)
\langle K^+\pi^-|Q_2^u|B^0_d\rangle_{\rm T}^{\rm fact}}\nonumber\\
&&~~~~\equiv\langle K^+\pi^-|Q_2^u(\mu)|B^0_d\rangle_{\rm T}+\left\{
\langle K^+\pi^-|Q_2^u(\mu)|B^0_d\rangle_{\rm P}-
\langle K^+\pi^-|Q_2^d(\mu)|B^0_d\rangle\right\},\label{B2-def}
\end{eqnarray}
we get
\begin{equation}\label{eps-r-ratio}
\frac{\epsilon}{r}\,e^{i(\Delta-\delta)}=-\,\frac{3}{2\lambda^2R_b}\left[
\frac{C_9(\mu)B_1(\mu)+3\,C_{10}(\mu)B_2(\mu)}{C_1'(\mu)B_1(\mu)+
3\,C_2'(\mu)B_2(\mu)}\right],
\end{equation}
where also the tiny electroweak penguin contributions to $T$, which have been
neglected for simplicity in (\ref{T-opex}), are included through
\begin{equation}
C_1'(\mu)\equiv C_1(\mu)+\frac{3}{2}\,C_9(\mu)\,,\quad
C_2'(\mu)\equiv C_2(\mu)+\frac{3}{2}\,C_{10}(\mu)\,.
\end{equation}

The expression (\ref{eps-r-ratio}) depends only on a single non-perturbative
parameter given by $B_2(\mu)/B_1(\mu)$, which is in general a complex
quantity owing to final-state interactions. It is possible to rewrite 
(\ref{eps-r-ratio}) in an even more transparent way by using the quantities
\begin{eqnarray}
a_1^{\rm eff}&\equiv&\frac{1}{3}\,C_1'(\mu)\,B_1(\mu)+C_2'(\mu)\,B_2(\mu)\\
a_2^{\rm eff}&\equiv&C_1'(\mu)\,B_2(\mu)+\frac{1}{3}\,C_2'(\mu)\,B_1(\mu)\,,
\end{eqnarray}
which correspond to the usual phenomenological colour factors $a_1$ and 
$a_2$ describing the intrinsic strength of colour-suppressed and 
colour-allowed decay processes, respectively \cite{ns}. Comparing 
experimental data on $B^-\to D^{(\ast)0}\pi^-$ and $\overline{B^0_d}\to 
D^{(\ast)+}\pi^-$, as well as on $B^-\to D^{(\ast)0}\rho^-$ and 
$\overline{B^0_d}\to D^{(\ast)+}\rho^-$ decays gives 
$a_2/a_1=0.26\pm0.05\pm0.09$ \cite{browder}, where $a_1$ and $a_2$ are
real quantities and their relative sign is interestingly found to be 
positive, which is in contrast to the case of $D$ decays.

A straightforward calculation yields
\begin{equation}
B(\mu)\equiv\frac{B_2(\mu)}{B_1(\mu)}=\frac{1}{3}\left[\frac{a\,e^{i\omega}
\,C_1(\mu)-C_2(\mu)}{C_1(\mu)-a\,e^{i\omega}\,C_2(\mu)}\right],
\end{equation}
where 
\begin{equation}
a\,e^{i\omega}\equiv\frac{a_2^{\rm eff}}{a_1^{\rm eff}}
\end{equation}
is in general 
also a complex quantity ($a=|a_2^{\rm eff}|/|a_1^{\rm eff}|$ is, however, 
real and positive). Inserting this expression for $B(\mu)$ into 
(\ref{eps-r-ratio}), we obtain
\begin{equation}
\frac{\epsilon}{r}\,e^{i(\Delta-\delta)}=\frac{3}{2\lambda^2R_b}\left[
\frac{C_1'(\mu)C_9(\mu)-C_2'(\mu)C_{10}(\mu)}{C_2'^2(\mu)-C_1'^2(\mu)}+
a\,e^{i\omega}\left\{\frac{C_1'(\mu)C_{10}(\mu)-C_2'(\mu)C_9(\mu)}{C_2'^2(\mu)-
C_1'^2(\mu)}\right\}\right].
\end{equation}
Since there is a strong cancellation in the first term, leading to 
\begin{equation}
\frac{C_1'(\mu)C_9(\mu)-C_2'(\mu)C_{10}(\mu)}{C_1'(\mu)C_{10}(\mu)-
C_2'(\mu)C_9(\mu)}={\cal O}(10^{-2})\,,
\end{equation}
we finally arrive at
\begin{equation}\label{eps-r-final}
\frac{\epsilon}{r}\,e^{i(\Delta-\delta)}\approx\frac{3}{2\lambda^2R_b}\left[
\frac{C_1'(\mu)C_{10}(\mu)-C_2'(\mu)C_9(\mu)}{C_2'^2(\mu)-
C_1'^2(\mu)}\right]a\,e^{i\omega} \approx 0.75\times a\,e^{i\omega}\,.
\end{equation}
The combination of Wilson coefficients in this expressions is essentially
renormalization-scale-independent and changes only by ${\cal O}(1\%)$ when
evolving from $\mu=M_W$ down to $\mu=m_b$. Employing $R_b=0.36$ and the 
leading-order Wilson coefficients 
\begin{equation}
C_1(m_b)=-\,0.308,\,\,\,C_2(m_b)=1.144,\,\,\,C_9(m_b)/\alpha=-\,1.280,
\,\,\,C_{10}(m_b)/\alpha=0.328
\end{equation}
obtained for $\Lambda_{\overline{{\rm MS}}}=225\,\mbox{MeV}$, we get the 
numerical value of 0.75 in (\ref{eps-r-final}), which we will apply throughout 
this section. The use of next-to-leading order Wilson coefficients would
change this result by only a few per cent. In view of the approximations made 
to derive (\ref{eps-r-final}) -- neglect of contributions from $Q_7$, $Q_8$
and electroweak penguins with internal charm quarks --  it is therefore 
appropriate to use leading-order Wilson coefficients. 

The ``factorization'' approach (corresponding to 
$B(\mu_{\rm F})=1$ in (\ref{eps-r-ratio}), where $\mu_{\rm F}$ is the 
``factorization scale'') gives on the other hand
\begin{equation}
\left.\frac{\epsilon}{r}\,e^{i(\Delta-\delta)}\right|_{\rm fact}=\,0.06\,,
\end{equation}
which is smaller than (\ref{eps-r-final}) for $a\,e^{i\omega}=0.25$ 
(corresponding to $B(m_b)\approx2/3$) by a factor of~3. 
For $\left.r\right|_{\rm fact}=0.16$ (see (\ref{r-fact})), we would have 
$\left.\epsilon\right|_{\rm fact}=0.01$, which is in accordance with the 
estimate given in \cite{fm2}.

The expression (\ref{eps-r-final}) for the electroweak penguin contribution 
to the $B\to\pi K$ amplitude relations (\ref{ampl-neut}) and (\ref{ampl-char})
shows that the usual terminology of ``colour-suppressed'' electroweak penguins
in this context is justified, since $P_{\rm ew}$ is proportional to the  
generalized ``colour factor'' $a_2^{\rm eff}$.

\boldmath
\subsection{The Decay $B^\pm\to\pi^\pm\pi^0$: A First Step Towards 
Constraining the Electroweak Penguin Uncertainty}\label{EWP-app-det}
\unboldmath
As we have noted above, a comparison of colour-suppressed and colour-allowed
$B\to D^{(\ast)}\pi(\rho)$ decays shows that the corresponding value of
$a_2/a_1$ is positive and of ${\cal O}(0.25)$. In the case of $B$-meson
decays into $\pi\,K$ or $\pi\,\pi$ final states, the situation concerning
``colour-suppression'' may, however, be quite different. At present there are
unfortunately no experimental data available to investigate this issue. A first
step towards achieving this goal, thereby obtaining insights into the 
importance of the colour-suppressed electroweak penguin amplitude $P_{\rm ew}$
in (\ref{ampl-char}), is provided by the $\bar b\to\bar u\,u\,\bar d$ decay 
$B^+\to\pi^+\pi^0$. This mode receives only tiny electroweak penguin 
contributions \cite{rev}. Moreover, QCD penguins do not contribute because 
of the $SU(2)$ isospin symmetry, so that the decay amplitude takes 
the simple form
\begin{equation}
A(B^+\to\pi^+\pi^0)=-\,\frac{1}{\sqrt{2}}\left[T^{(d)}+C^{(d)}\right]=
-\,\frac{e^{i\gamma}}{\sqrt{2}}\left[|T^{(d)}|\,e^{i\delta_T^{(d)}}+
|C^{(d)}|\,e^{i\delta_C^{(d)}}\right],
\end{equation}
where $T^{(d)}$ and $C^{(d)}$ are usually referred to as colour-allowed
and colour-suppressed ``tree'' amplitudes. Using the $SU(3)$ flavour symmetry,
we have 
\begin{equation}
|T^{(d)}|=\frac{f_\pi}{\lambda\,f_K}\,\lambda^4A\,R_b\,|\tilde {\cal T}|\,,
\end{equation}
where $\tilde {\cal T}$ has been introduced in (\ref{T-def}), and $f_\pi$ and 
$f_K$ denote the pion and kaon decay constants, respectively, taking into 
account factorizable $SU(3)$ breaking. Introducing
\begin{equation}
a_{\pi\pi}\,e^{i\,\omega_{\pi\pi}}\equiv\frac{C^{(d)}}{T^{(d)}}\,,
\end{equation}  
we find
\begin{equation}\label{apipi}
1+2\,a_{\pi\pi}\,\cos\omega_{\pi\pi}+a_{\pi\pi}^2\,=\,\left(\frac{1}{\tilde r}
\,\frac{\lambda\,f_K}{f_\pi}\right)^2
\frac{2\,\mbox{BR}(B^\pm\to\pi^\pm\pi^0)}{\mbox{BR}
(B^\pm\to\pi^\pm K)}
\end{equation}
and get a lower bound on $a_{\pi\pi}$:
\begin{equation}\label{bound-apipi}
a_{\pi\pi}\,\ge\,\left|\,1-\frac{1}{\tilde r}\,\frac{\lambda\,f_K}{f_\pi}\,
\sqrt{\frac{2\,\mbox{BR}(B^\pm\to\pi^\pm\pi^0)}{\mbox{BR}
(B^\pm\to\pi^\pm K)}}\,\right|,
\end{equation}
and an upper limit for $|\sin\omega_{\pi\pi}|$:
\begin{equation}\label{bound-omegapipi}
|\sin\omega_{\pi\pi}|\,\le\,\frac{1}{\tilde r}\,\frac{\lambda\,f_K}{f_\pi}\,
\sqrt{\frac{2\,\mbox{BR}(B^\pm\to\pi^\pm\pi^0)}{\mbox{BR}
(B^\pm\to\pi^\pm K)}}\,,
\end{equation}
where $\tilde r$ corresponds to $\tilde {\cal T}$ and has been defined in
(\ref{Def-r-tilde}). At present there is only an upper bound on the 
combined branching ratio for $B^\pm\to\pi^\pm\pi^0$ available from CLEO
\cite{cleo}, which is given by BR$(B^\pm\to\pi^\pm\pi^0)<2.0\times10^{-5}$ 
and is unfortunately too weak to constrain $a_{\pi\pi}$ and $\omega_{\pi\pi}$
in a meaningful way.

Making use once more of the $SU(3)$ flavour symmetry, we get
\begin{eqnarray}
\lefteqn{a\,e^{i\omega}\approx}\label{a-SU3rel}\\
&&\frac{a_2^{\pi\pi}+\left[\sum\limits_{k=1}^2C_k(\mu)\left\{
\langle K^+\pi^-|Q_{3-k}^u(\mu)|B^0_d\rangle_{\rm P}-
\langle K^+\pi^-|Q_{3-k}^d(\mu)|B^0_d\rangle\right\}\right]/
\langle K^+\pi^-|Q_2^u|B^0_d\rangle_{\rm T}^{\rm fact}}
{a_1^{\pi\pi}+\left[\sum\limits_{k=1}^2C_k(\mu)\left\{
\langle K^+\pi^-|Q_k^u(\mu)|B^0_d\rangle_{\rm P}-
\langle K^+\pi^-|Q_k^d(\mu)|B^0_d\rangle\right\}\right]/
\langle K^+\pi^-|Q_2^u|B^0_d
\rangle_{\rm T}^{\rm fact}}\nonumber
\end{eqnarray}
with $a_2^{\pi\pi}/a_1^{\pi\pi}=a_{\pi\pi}\,e^{i\omega_{\pi\pi}}$. This 
equation can easily be rewritten as
\begin{equation}\label{a-SU3relsimpl}
a\,e^{i\omega}\approx\left(\frac{1+a_2^{\rm res}/a_2^{\pi\pi}\,
{\cal M}_{\rm res}}{1+{\cal M}_{\rm res}}\right)a_{\pi\pi}\,
e^{i\omega_{\pi\pi}},
\end{equation}
where we have expressed the terms in the denominator and
numerator of (\ref{a-SU3rel}) related to rescattering processes as 
$a_1^{\pi\pi}{\cal M}_{\rm res}$ and $a_2^{\rm res}{\cal M}_{\rm res}$, 
respectively. In general these rescattering contributions preclude a relation
of $a_{\pi\pi}\,e^{i\omega_{\pi\pi}}$ to $a\,e^{i\omega}$. However, the
rescattering contributions from the $Q_{1}^{u,d}$ current--current operators
are disfavoured with respect to those from $Q_{2}^{u,d}$ because of their
colour structure. This feature is described by the colour-suppression 
factor $a_2^{\rm res}$ in (\ref{a-SU3relsimpl}). If this quantity should 
have the same order of magnitude as $a_2^{\pi\pi}$, we would have 
$a\,e^{i\omega}\approx a_{\pi\pi}\,e^{i\omega_{\pi\pi}}$ not only in the 
case of tiny rescattering effects, i.e.\ $|{\cal M}_{\rm res}|\ll1$, but 
also for large rescattering contributions. It is interesting to note that 
we would have $a\,e^{i\omega}\approx C_1(m_b)/C_2(m_b)\approx-\,0.25$, 
if the rescattering processes from $Q_{2}^{u,d}$ should play a dominant role. 

Combining these considerations, we conclude that BR$(B^\pm\to\pi^\pm\pi^0)$ 
is interesting to obtain a lower bound on the electroweak penguin 
contributions (see (\ref{eps-r-final}) and (\ref{bound-apipi})), or to 
eliminate either $\omega$ or $a$ (see (\ref{apipi})). This strategy -- using 
ironically $B^+\to\pi^+\pi^0$, a decay where electroweak penguins play a 
very minor role -- can be considered as a first step towards constraining 
the electroweak penguin amplitude $P_{\rm ew}$ in the $B\to\pi K$ 
relations (\ref{ampl-neut}) and (\ref{ampl-char}). Its theoretical accuracy 
is limited by $SU(3)$-breaking corrections, which may be significant in the 
case of colour-suppressed topologies, and by rescattering effects. In the 
numerical examples given in the following subsection, we assume 
$a={\cal O}(0.25)$, i.e.\ that the $B\to D^{(\ast)}\pi(\rho)$ and 
$B\to J/\psi\,K^{(\ast)}$ measurements available at present \cite{browder}
inform us also about the intrinsic ``strength'' of colour suppression in 
$B\to\pi K$ decays, and keep $\omega$ as a free parameter.

\boldmath
\subsection{A Closer Look at Electroweak Penguin Effects in Strategies 
to Constrain and Determine $\gamma$}\label{EWP-effects2}
\begin{figure}
\centerline{
\rotate[r]{
\epsfxsize=9.2truecm
\epsffile{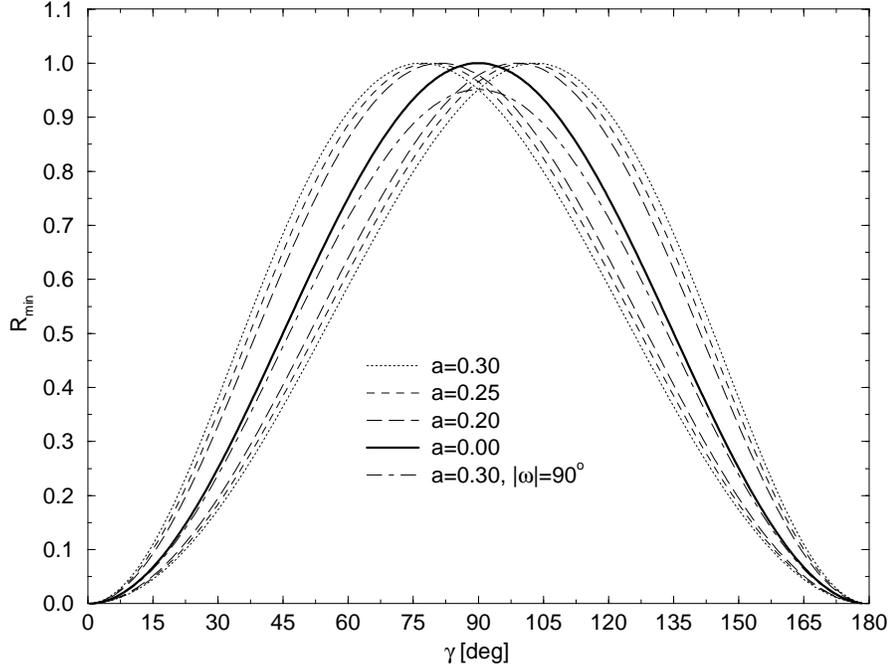}}}
\caption{The effect of electroweak penguins on $R_{\rm min}$ described by 
(\ref{RminEWP2}). The curves for a given value of $a$ (except the dot-dashed 
line) correspond to $\omega\in\{0^\circ,180^\circ\}$.}\label{fig:Rmin-a}
\end{figure}

\unboldmath
Since (\ref{eps-r-final}) implies a correlation between $\epsilon$ and $r$
described by
\begin{equation}\label{eps-r-rel}
\epsilon=q\,r\,,\quad \Delta=\delta+\omega\,,
\end{equation}
where $q\approx0.75\times a$, (\ref{R-simpl}) is modified. Using 
(\ref{eps-r-rel}) to replace $\epsilon$ and $\Delta$ in (\ref{R-exp}), we get
\begin{equation}\label{R-exp2}
R=1-\frac{2\,r}{w}\left(\tilde h\cos\delta+\tilde k\sin\delta\right)+v^2r^2,
\end{equation}
with
\begin{equation}
v=\sqrt{1-2\,q\,\cos\omega\cos\gamma+q^2}
\end{equation}
and
\begin{eqnarray}
\tilde h&=&\cos\gamma+\rho\cos\theta-q\left[\,\cos\omega+
\rho\cos(\theta-\omega)\cos\gamma\,\right]\\
\tilde k&=&\rho\sin\theta+q\left[\,\sin\omega-\rho\sin(\theta-\omega)
\cos\gamma\,\right].
\end{eqnarray}
If we keep both $r$ and the strong phase $\delta$ as free parameters in
(\ref{R-exp2}), we find that $R$ takes the following minimal value:
\begin{equation}\label{RminEWP2}
R_{\rm min}=\left[\frac{1+2\,q\,\rho\,\cos(\theta+\omega)+
q^2\rho^2}{\left(1-2\,q\,\cos\omega\cos\gamma+q^2\right)
\left(1+2\,\rho\,\cos\theta\cos\gamma+\rho^2\right)}\right]\sin^2\gamma\,,
\end{equation}
which simplifies to $R_{\rm min}=(\sin^2\gamma)/(1-
2\,q\,\cos\omega\cos\gamma+q^2)$ for $\rho=0$, i.e.\ in the case of neglected 
rescattering effects. In Fig.~\ref{fig:Rmin-a} we have illustrated the latter 
expression for various values of $a$. The curves shifted to the left correspond
to $\omega=0^\circ$, while those shifted to the right correspond to 
$\omega=180^\circ$ and represent the maximal electroweak penguin effects for 
a given value of $a$. For $|\omega|=90^\circ$, these effects are minimal and 
only of second order in $a$.

As was pointed out in Subsection~\ref{gam-bounds}, the strong phase
$\delta$ in (\ref{R-exp2}) can be eliminated with the help of the
pseudo-asymmetry $A_0$. The modification of (\ref{A-Def}) through 
(\ref{eps-r-rel}) is as follows:
\begin{equation}\label{det-delta2}
\tilde A=\tilde B\sin\delta-\tilde C\cos\delta\,,
\end{equation}
where
\begin{equation}
\tilde A=\frac{w}{r}\left[\,\frac{A_0-A_+}{2\sin\gamma}+q\,
r^2\sin\omega\,\right],\,\,\,\,
\tilde B=1+q\,\rho\,\cos(\theta-\omega)\,,\,\,\,\,
\tilde C=q\,\rho\,\sin(\theta-\omega)\,,
\end{equation}
and gives
\begin{equation}\label{R-simpl2}
R=L+M\,r^2\mp\sqrt{-\,N\,r^4+2\,P\,r^2-Q}\,,
\end{equation}
with
\begin{displaymath}
L=1-\left(\frac{A_0-A_+}{2\sin\gamma}\right)\tilde D\,,\,\,\,\,
M=v^2-\tilde D\,q\,\sin\omega\,,\,\,\,\,
N=(\tilde E\,q\,\sin\omega)^2
\end{displaymath}
\begin{equation}
P=\left[\,\frac{\tilde B^2+\tilde C^2}{2\,w^2}-\left(\frac{A_0-A_+}{2\,
\sin\gamma}\right)\,q\,\sin\omega\,\right]\tilde E^2,\,\,\,\,
Q=\left[\left(\frac{A_0-A_+}{2\sin\gamma}\right)\tilde E\right]^2
\end{equation}
and
\begin{equation}
\tilde D=2\left(\frac{\tilde k\,\tilde B-
\tilde h\,\tilde C}{\tilde B^2+\tilde C^2}\right),\,\,\,\,
\tilde E=2\left(\frac{\tilde h\,\tilde B+
\tilde k\,\tilde C}{\tilde B^2+\tilde C^2}\right).
\end{equation}
If we keep $r$ as a free parameter in (\ref{R-simpl2}), we find that $R$
takes minimal and maximal values, which are given by
\begin{equation}\label{Rmin-compl}
R_{\rm min}^{\rm max}=\frac{1}{N}\left[\,LN+MP\pm\sqrt{(P^2-NQ)
(M^2+N)}\,\right]
\end{equation}
and correspond to 
\begin{equation}
r=r_{\rm min}^{\rm max}\equiv\sqrt{\frac{P}{N}\pm\frac{M}{N}
\sqrt{\frac{P^2-NQ}{M^2+N}}}\,.
\end{equation}
Moreover we get the following expression for $r$:
\begin{equation}\label{r-expr-a}
r=\sqrt{r_{\rm ext}^2+\frac{(R-R_{\rm ext})M\pm\sqrt{(R-R_{\rm ext})
\left[\,2\,MP+(2\,L-R-R_{\rm ext})N\,\right]}}{M^2+N}}\,,
\end{equation}
where ``ext'' stands for either ``min'' or ``max'', i.e.\ denotes the
extremal values. It is possible to rewrite (\ref{Rmin-compl}) in a more
transparent way as follows:
\begin{equation}\label{Rmin-a-expr}
R_{\rm min}^{\rm max}=\frac{1}{2\,(w\,q\,\sin\omega)^2}\left[\,v^2y\pm
\sqrt{(y^2-z^2)(v^4-4\,q^2\sin^2\omega\,\sin^2\gamma)}\,\right],
\end{equation}
where
\begin{equation}
y=1+2\,q\,\rho\,\cos(\theta+\omega)+q^2\rho^2-z\,,\quad
z=\left(\frac{w^2A_0}{\sin\gamma}\right)q\,\sin\omega\,.
\end{equation}
Concerning phenomenological applications, only the minimal value of $R$
plays an important role, since $R^{\rm max}$ turns out to be much larger 
than 1. 

In the case of two interesting special cases, (\ref{Rmin-a-expr}) simplifies 
considerably. First, for $A_0=0$ we have
\begin{equation}
\left.R_{\rm min}\right|_{A_0=0}=
\frac{2}{v^2w^2}\left[\frac{1+2\,q\,\rho\cos(\theta+\omega)+
q^2\rho^2}{1+
\sqrt{1-\left(2\,q\,\sin\omega\,\sin\gamma/v^2\right)^2}}\right]\sin^2\gamma\,,
\end{equation}
which agrees with (\ref{RminEWP2}) for $q\,\sin\omega=0$, and
gives larger values for $R_{\rm min}$, i.e.\ stronger bounds on $\gamma$, 
otherwise. Another important case is $q\,\sin\omega=0$, corresponding to 
$\omega\in\{0^\circ,180^\circ\}$ for $q\not=0$. In this case, $R_{\rm min}$
takes the same form as (\ref{Rmin}). The expression for $\kappa$ is, however, 
very different (and $\cos\omega=\pm1$):
\begin{equation}
\kappa=\frac{1+2\,q\,\rho\,\cos\theta\,\cos\omega+
q^2\rho^2}{(1-2\,q\,\cos\omega\,\cos\gamma+q^2)(1+2\,\rho\,\cos\theta\,
\cos\gamma+\rho^2)}\,.
\end{equation}

\begin{figure}
\centerline{
\rotate[r]{
\epsfxsize=9.2truecm
\epsffile{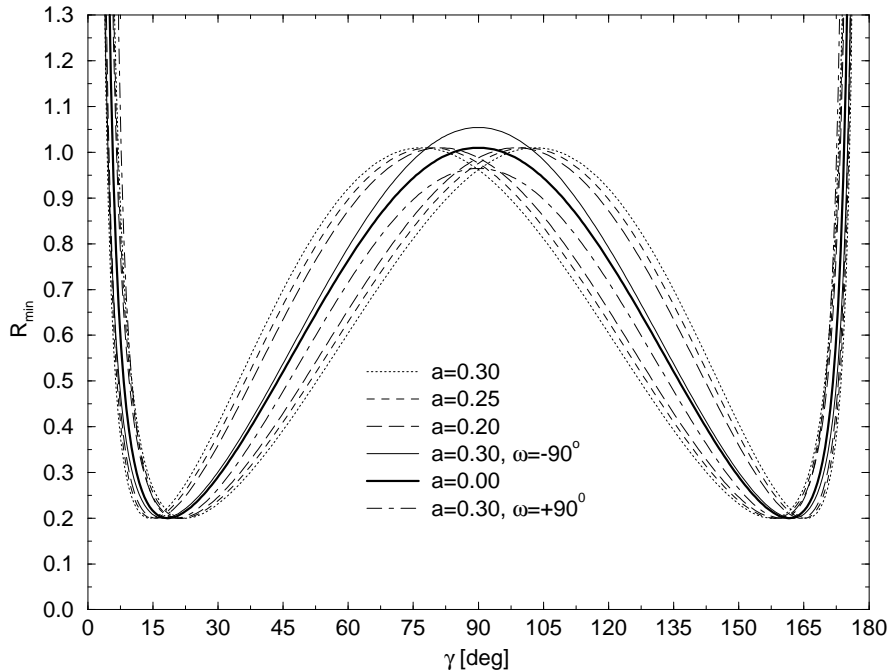}}}
\caption{The effect of electroweak penguins described by (\ref{eps-r-final})
on $R_{\rm min}$ for $A_0=+\,0.2$. Except for the solid and dot-dashed lines, 
the curves correspond to 
$\omega\in\{0^\circ,180^\circ\}$.}\label{fig:Rmin-aA02}
\end{figure}

\begin{figure}
\centerline{
\rotate[r]{
\epsfxsize=9.2truecm
\epsffile{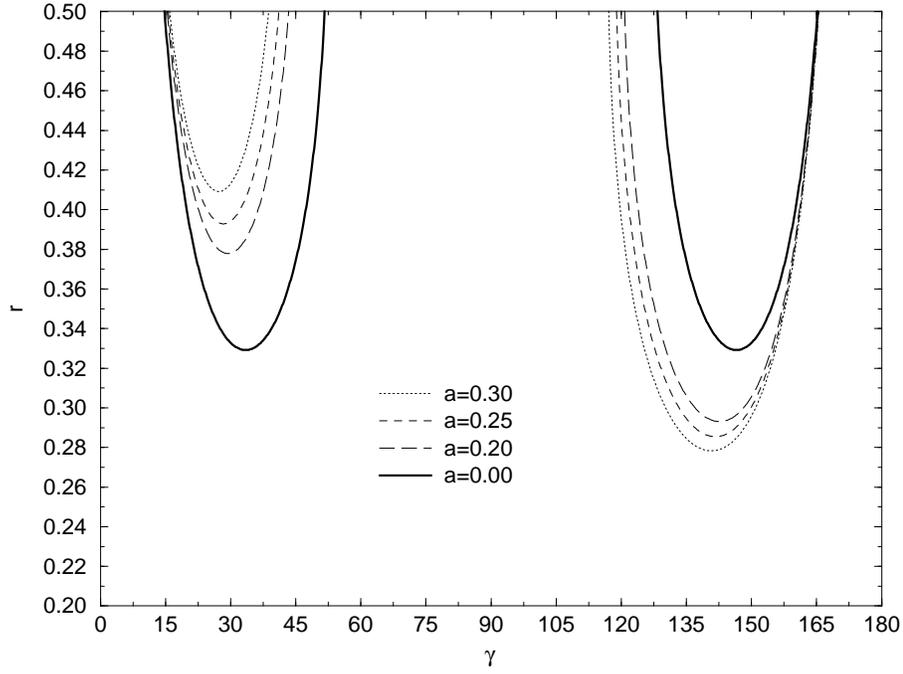}}}
\caption{The shift of the contours in the $\gamma$--$r$ plane corresponding
to $R=0.65$, $|A_0|=0.2$ through electroweak penguins described by 
(\ref{eps-r-final}) for $\omega=0^\circ$ and 
$\rho=0$.}\label{fig:r-det065ew-a0}
\end{figure}

\begin{figure}
\centerline{
\rotate[r]{
\epsfxsize=9.2truecm
\epsffile{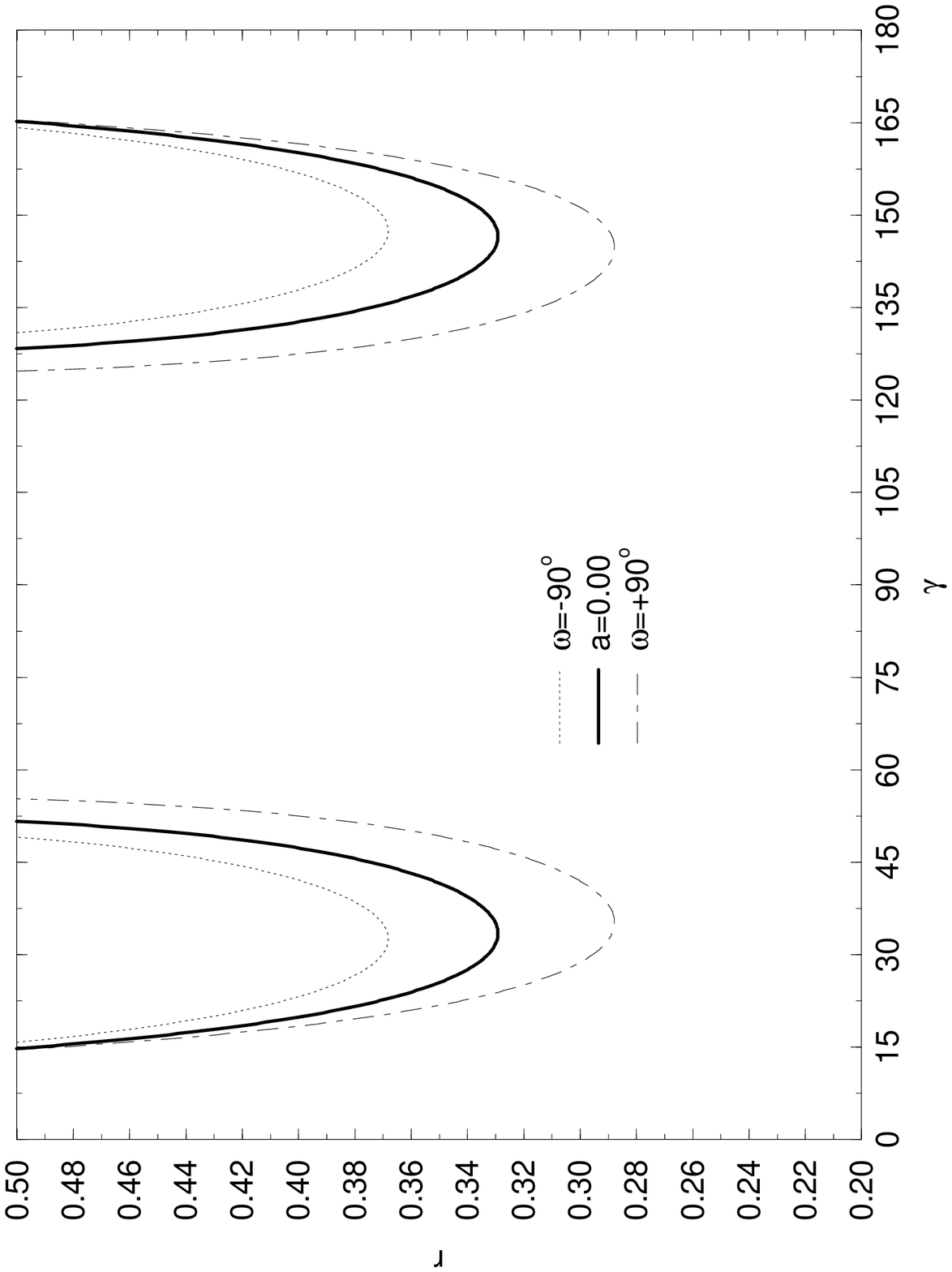}}}
\caption{The shift of the contours in the $\gamma$--$r$ plane corresponding
to $R=0.65$, $A_0=0.2$ through electroweak penguins described by 
(\ref{eps-r-final}) for $a=0.3$ and $\rho=0$.}\label{fig:r-det065ew-a90}
\end{figure}

In Fig.~\ref{fig:Rmin-aA02} we illustrate the effect of electroweak penguins
described by (\ref{eps-r-final}) on $R_{\rm min}$ for $A_0=+\,0.2$ by using 
(\ref{Rmin-a-expr}) for various values of $a$ and neglected rescattering 
effects, i.e.\ $\rho=0$. The curves shifted to the left correspond to 
$\omega=0^\circ$, those shifted to the right to $\omega=180^\circ$.
In Figs.~\ref{fig:r-det065ew-a0} and \ref{fig:r-det065ew-a90} we show the 
corresponding effects in the $\gamma$--$r$ plane for $\omega=0^\circ$ and
$\pm90^\circ$, respectively. In the latter case, we have chosen $a=0.3$.
The contours shown in these figures have been calculated by using
(\ref{r-expr-a}). The case corresponding to $\omega=180^\circ$ can 
easily be obtained from Fig.~\ref{fig:r-det065ew-a0} by replacing 
$\gamma\to180^\circ-\gamma$. 

\begin{figure}
\centerline{
\rotate[r]{
\epsfxsize=9.2truecm
\epsffile{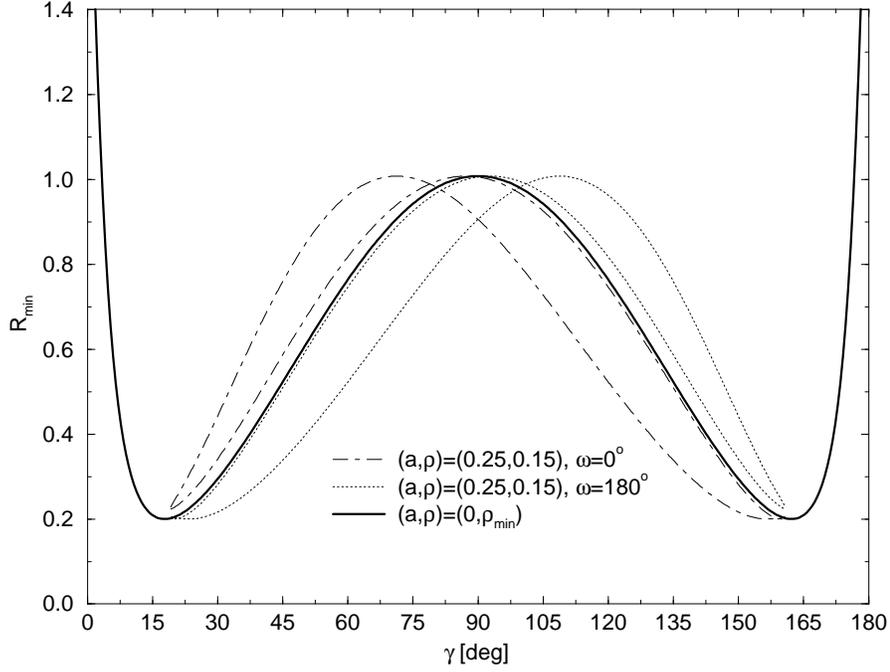}}}
\caption{Electroweak penguin and rescattering effects for $R_{\rm min}$ in the
case of $|A_0|=0.2$ and $|A_+|=0.1$ 
($\omega\in\{0^\circ,180^\circ\}$).}\label{fig:Rminres-a1.ps}
\end{figure}

\begin{figure}
\centerline{
\rotate[r]{
\epsfxsize=9.2truecm
\epsffile{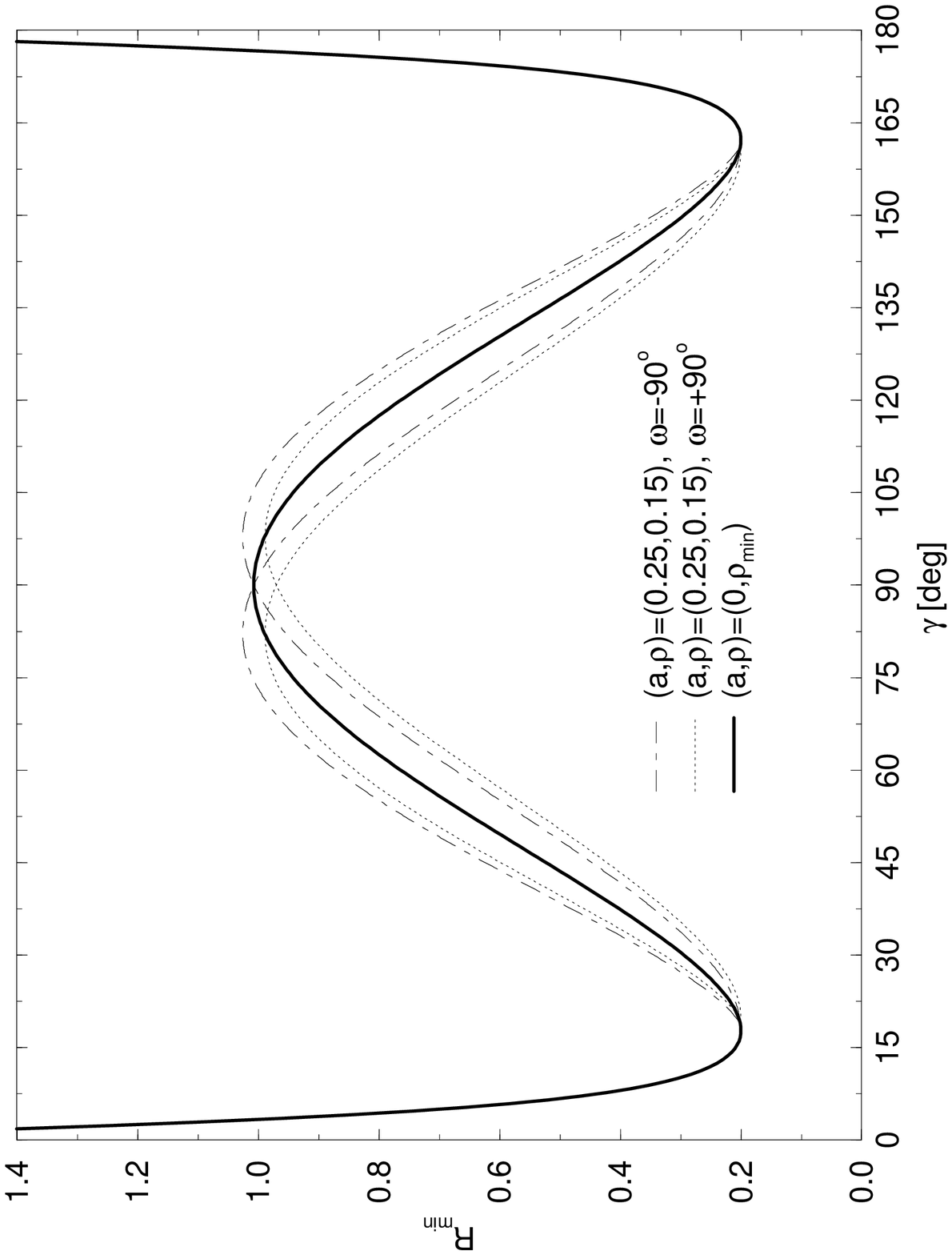}}}
\caption{Electroweak penguin and rescattering effects for $R_{\rm min}$ in the
case of $A_0=0.2$ and $|A_+|=0.1$ 
($\omega=\pm90^\circ$).}\label{fig:Rminres-a2}
\end{figure}

\boldmath
\section{Combined Rescattering and Electroweak Penguin Effects}\label{comb-eff}
\unboldmath
To end the discussion of rescattering and electroweak penguin effects 
in strategies to constrain and determine the CKM angle $\gamma$ from 
$B^\pm\to\pi^\pm K$ and $B_d\to\pi^\mp K^\pm$ decays, let us illustrate
the situation concerning $R_{\rm min}$ in the case of measured asymmetries
$|A_0|=0.2$ and $|A_+|=0.1$. The latter CP asymmetry allows us to obtain
a lower bound on $\rho$ and to eliminate the CP-conserving strong phase 
$\theta$ for a given value of $\rho$, as we discussed in 
Section~\ref{FSI-effects}. In Figs.~\ref{fig:Rminres-a1.ps} and 
\ref{fig:Rminres-a2} we have chosen $\rho=0.15$. The electroweak penguin
effects are described in these figures by $a=0.25$ and various values of
the strong phase $\omega$. We see that an important difference arises 
between $\omega=0^\circ$ and $\omega=180^\circ$.

In the case of large rescattering effects, for example $\rho=0.15$, as shown 
in Figs.~\ref{fig:Rminres-a1.ps} and \ref{fig:Rminres-a2}, the branching ratio
for the decay $B^+\to K^+\overline{K^0}$ may be enhanced by one order of 
magnitude from its ``short-distance'' value ${\cal O}(10^{-6})$ to the 
$10^{-5}$ level, which should be accessible at future $B$ factories.
Using the $SU(3)$ flavour symmetry to relate this mode to $B^+\to\pi^+
K^0$, the rescattering effects affecting $R_{\rm min}$ can be controlled 
completely, as we saw in Subsection~\ref{FSI-incl}. In particular, following
this strategy, no knowledge about $\rho$ would be needed as it was in
Figs.~\ref{fig:Rminres-a1.ps} and \ref{fig:Rminres-a2}. 

Although we could derive a transparent expression (see (\ref{eps-r-final})) 
to describe the electroweak penguin contributions affecting the isospin 
relations (\ref{ampl-neut}) and (\ref{ampl-char}) between $B^+\to\pi^+ K^0$ 
and $B^0_d\to\pi^-K^+$, it is more difficult to control them using 
experimental data if rescattering effects are large. A first step in this 
direction is provided by the branching ratio for $B^+\to\pi^+\pi^0$, 
as we have pointed out in Subsection~\ref{EWP-app-det}. In 
Figs.~\ref{fig:Rminres-a1.ps} and~\ref{fig:Rminres-a2} we have assumed that
$a$ takes a value of $0.25$, which is of the same order of magnitude as
the strength of colour suppression in $B\to D^{(\ast)}\pi(\rho)$ and 
$B\to J/\psi\,K^{(\ast)}$ decays, and have kept $\omega$ as a free 
CP-conserving strong phase. 

At this point we could give more examples to illustrate the possible impact
of combined rescattering and electroweak penguin effects on information on
the CKM angle $\gamma$ obtained from $B\to\pi K$ decays. Since it is now
an easy exercise to play with the corresponding formulae, we will not 
consider other scenarios in this paper. Hopefully, improved experimental 
data will be available in the near future, allowing us to go beyond these 
selected examples and to perform a solid analysis of the corresponding 
decays.

\section{Conclusions}\label{conclu}
In summary, we have presented a general parametrization of the 
$B^+\to\pi^+ K^0$ and $B^0_d\to\pi^-K^+$ decay amplitudes within the 
framework of the Standard Model in terms of ``physical quantities'', taking 
into account both rescattering and electroweak penguin effects. These decays 
offer an experimentally feasible way to obtain direct information on the CKM 
angle $\gamma$ at planned $B$ factories, which will start operating in the 
near future, and are therefore of particular phenomenological interest. In 
this respect, the ratio $R$ of their combined branching ratios and 
the pseudo-asymmetry $A_0$ play the key role. As soon as these 
observables, i.e.\ the branching ratios for $B^+\to\pi^+ K^0$, 
$B^0_d\to\pi^-K^+$ and their charge-conjugates, have been measured, contours 
in the $\gamma$--$r$ plane can be calculated with the help of the formulae 
derived in this paper. These contours imply allowed ranges for both $r$ and 
$\gamma$. For $A_0\not=0$, values of $\gamma$ within intervals around 
$0^\circ$ and $180^\circ$ can be excluded, and if $R$ should turn out to 
be smaller than 1, also values around $90^\circ$ can be ruled out. In 
particular the latter case is of particular interest, since the corresponding 
range for $\gamma$ would then be complementary to its presently allowed 
range obtained from the usual fits of the unitarity triangle. If $r$ could 
be fixed by using an additional input, $\gamma$ could not only be 
constrained, but determined up to a four-fold ambiguity.

In order to derive these bounds and to obtain the contours in the 
$\gamma$--$r$ plane, isospin symmetry has been used, which is certainly 
an excellent working assumption. The theoretical cleanliness is, however, 
limited by certain rescattering and electroweak penguin effects. Our formulae 
include these contributions in a completely general way, and therefore 
allow us to investigate the sensitivity to these effects and to take them
into account by using additional experimental information. 

The rescattering processes may lead to sizeable CP violation in 
$B^+\to\pi^+ K^0$, possibly as large as ${\cal O}(10\%)$. We have pointed 
out that such CP asymmetries would provide a lower bound on $\rho$, i.e.\ 
a first constraint on the strength of the rescattering effects, and are 
useful to include these final-state interactions in the bounds on 
$\gamma$. In order to control the rescattering effects in these bounds,
$B^+\to K^+\overline{K^0}$ may be a ``gold-plated'' mode. It can be 
related to $B^+\to \pi^+ K^0$ with the help of the $SU(3)$ flavour symmetry 
and provides sufficient information not just to constrain, but to take into 
account the final-state interaction effects in the bounds on $\gamma$
completely. As a by-product, this strategy gives moreover an allowed region 
for $\rho$, and excludes values of $\gamma$ within ranges around $0^\circ$ 
and $180^\circ$. It is interesting to note that $SU(3)$ breaking enters in
this approach only at the ``next-to-leading order'' level, as it represents a 
correction to the correction to the bounds on $\gamma$ arising from 
rescattering processes. Moreover this strategy works also if the CP 
asymmetry in $B^+\to\pi^+ K^0$ should turn out to be very small. In this 
case there may also be large rescattering effects, which would then not be 
signalled by sizeable CP violation in this channel.

At first sight, an experimental study of $B^+\to K^+ \overline{K^0}$ appears 
to be challenging, since model estimates performed at the perturbative 
quark level give a combined branching ratio 
BR$(B^\pm\to K^\pm K)={\cal O}(10^{-6})$, which is one order of magnitude 
below the present upper limit obtained by the CLEO collaboration. However, 
as we have pointed out in this paper, rescattering processes may well enhance 
this branching ratio by ${\cal O}(10)$, so that it may actually be much closer
to the CLEO limit than na\"\i vely expected. Consequently, in the case of 
large rescattering contributions, i.e.\ when we have to take them into account 
in the bounds on $\gamma$, the branching ratio for the decay allowing us 
to accomplish this task, $B^+\to K^+ \overline{K^0}$, may be significantly
enhanced through these very rescattering effects, so that this strategy 
should be feasible at future $B$ factories. 

Although the decay $B^+\to K^+ \overline{K^0}$ allows us to determine the
shift of the contours in the $\gamma$--$r$ plane arising from rescattering
processes, it does not allow us to take into account these effects also in
the determination of $\gamma$, requiring some knowledge on $r$, in contrast 
to the $\gamma$ bounds. This quantity is not just the ratio of a ``tree'' to a
``penguin'' amplitude, which is the usual terminology in the literature, but 
has a rather complex structure and may be considerably affected by 
final-state interactions. Consequently, in the case of large rescattering 
effects, $r$ is expected to be shifted significantly from its ``factorized'' 
value and its theoretical uncertainty is very hard to control. Interestingly, 
the small present central value $R=0.65$ implies a range for $r$ that is at 
the edge of compatibility with these ``factorized'' results and favours larger 
values of $r$. This feature may already give us a first hint that 
rescattering effects play in fact an important role, and it may well be 
that future experimental results for $R$ will stabilize at the 0.65 level. 
In this case, a measurement of BR$(B^\pm\to K^\pm K)={\cal O}(10^{-5})$ and 
large CP violation in this channel would not be a surprise. In the recent 
literature it has been claimed by some authors that such rescattering 
processes would spoil the bounds on $\gamma$. It is interesting to note that 
these effects may actually be responsible for a strong realization of these 
bounds, i.e.\ for a small value of $R$, thereby making them 
phenomenologically relevant. 

Concerning electroweak penguins, model calculations using ``factorization'' 
to deal with hadronic matrix elements typically give contributions at the 
$1\%$ level in the case of $B^+\to\pi^+ K^0$ and $B^0_d\to\pi^-K^+$ decays, 
where electroweak penguins contribute only in ``colour-suppressed'' form. In 
this paper, we have presented an improved theoretical description of the 
electroweak penguin amplitude affecting the isospin relations between
the decay amplitudes of these modes, and have derived a transparent 
expression, clarifying also the notion of ``colour-suppressed'' electroweak 
penguins. Our approach does not use questionable assumptions, such as 
factorization, and makes use of only the general structure of the electroweak 
penguin operators and of the isospin symmetry of strong interactions to relate 
the hadronic matrix elements corresponding to $B^+\to\pi^+ K^0$ and 
$B^0_d\to\pi^-K^+$ transitions. We have seen that the importance of electroweak
penguins is closely related to the ratio of certain ``effective'' colour 
factors $a_2^{\rm eff}/a_1^{\rm eff}$. Using $|a_2^{\rm eff}|/|a_1^{\rm eff}|=
0.25$ gives an enhancement of the relevant electroweak penguin amplitude by a
factor of~3 with respect to the factorized result. A first step towards 
constraining this electroweak penguin amplitude experimentally is provided by 
the mode $B^+\to\pi^+\pi^0$. Our formulae include the electroweak penguin 
contributions in a completely general way, allowing us to take them into 
account once we have a better understanding of ``colour-suppression'' and 
rescattering effects in $B\to\pi K$ decays. 

Although the decays $B^+\to\pi^+ K^0$, $B^0_d\to\pi^-K^+$ and their 
charge conjugates will probably not allow a precision measurement of 
$\gamma$, they are expected to provide a very fertile ground to constrain 
this CKM angle. An accurate measurement of these modes, as well as of 
$B^\pm\to K^\pm K$, is therefore an important goal of the future $B$ 
factories. The corresponding experimental results will certainly be very 
exciting.

\end{document}